\documentclass[aps,prl,groupedaddress,superscriptaddress,twocolumn]{revtex4-2}

\usepackage{amsmath,graphicx}
\usepackage{float}
\usepackage{subfigure}
\usepackage{color}
\usepackage{braket}
\usepackage{mathrsfs}
\usepackage{appendix}
\usepackage{amssymb}
\usepackage{bm}
\usepackage{amsthm}
\usepackage{amsfonts}
\usepackage[T1]{fontenc}

\begin{document}
\title{Atomic collapse of high-order singular potentials in graphene}
\author{Yu-Chen Zhuang}
\affiliation{International Center for Quantum Materials, School of Physics, Peking University, Beijing 100871, China}

\author{Yue Mao}
\affiliation{International Center for Quantum Materials, School of Physics, Peking University, Beijing 100871, China}

\author{Qing-Feng Sun}
\email[]{sunqf@pku.edu.cn}
\affiliation{International Center for Quantum Materials, School of Physics, Peking University, Beijing 100871, China}
\affiliation{Hefei National Laboratory, Hefei 230088, China}

\date{\today}

\begin{abstract}
	Artificial atoms in graphene hosting a series of quasi-bound states can serve as an excellent platform to explore atomic collapse and become a basis to design novel graphene nanodevices. We theoretically study behaviors of massless Dirac fermions in singular potentials with a general form of $1/r^\gamma$. Different from the Coulomb potential that demands a supercritical charge $Z > Z_c$, a high-order singular potential ($\gamma > 1$) is found to in principle induce atomic collapse with an infinitesimal charge $Z$. The energies of atomic collapse states (ACSs) within these potentials are arranged roughly as a power sequence. We also show that some special ACSs can exist even above the bulk Dirac point, which cannot appear in the Coulomb potential. These findings uncover the anomalies of massless Dirac fermions in diverse charge potentials and provide guidance for further experiments and graphene nanodevice applications.
\end{abstract}

\maketitle

\textit{Introduction}. Due to massless Dirac fermionic excitations, graphene is regarded as an ideal two-dimensional system to simulate ultra-relativistic physics and quantum electrodynamics \cite{NovoselovNature2005,CastroRMP2009}.
One of the most attractive phenomena is atomic collapse \cite{Darwin}.
It means classical atomic orbits will be unstable and electrons will eventually collapse towards the nucleus once the nuclear charge is supercritical $Z > Z_c$. Since the predicted critical charge $Z_c \approx 137$ is very large \cite{Greiner}, the observation of atomic collapse in high energy physics experiments is still lacking.
However, due to a large fine-structure constant in graphene, just a small charge $Z \geq 1$ is sufficient to induce similar atomic collapse \cite{ShytovPRL2007}.
This has aroused broad interest to investigate Coulomb scatterings and electrical confinements in graphene \cite{PereiraPRL2007,WangScience2013,GeNano2021,GeNatN2023,GeNature2024,RenACS2025}.

The existence of Klein tunneling renders chiral massless fermions a high transmissibility to cross potential wells at normal incidence \cite{KatsnelsonNatP2006, LamasPRB2024}. Therefore, unlike two-dimensional electron gas based on semiconductors \cite{MittagPRX2021}, it is a challenge to control the motion of charge carriers in graphene \cite{LiFront2021}. However, atomic collapse offers a possibility to trap massless electrons for a finite time. In supercritical Coulomb potentials, a series of quasi-bound states will emerge below the bulk Dirac point, i.e., atomic collapse states (ACSs) to manifest collapsing atomic orbits towards the potential center \cite{ShytovPRL2007, SilvaPRB2023, WangPRB2024}. Like Efimov states \cite{Efimov}, the energies of ACSs are arranged in geometric sequences. It reflects a scale anomaly where the continuous scale symmetry of the system breaks into discrete scale invariance \cite{OvdatNC2017,NishidaPRB2014,NishidaPRB2016,ShaoPNAS2022}. Experiments have developed many methods to introduce effective local electric potentials and implement quasi-bound states, such as introducing charged impurities \cite{WangScience2013}, charged defects \cite{MaoNP2016}, decorated scanning tunneling microscopy (STM) tips \cite{JiangNN2017}, and quantum dots \cite{Zheng,Zheng2,MaoNature2025}, etc. By using techniques like voltage pulses and dual-gate control \cite{MaoNP2016,JiangNN2017}, experiments are able to adjust the strength and distribution of these local potentials and transform the originally subcritical potential into the supercritical potential along with a series of ACSs. The exploration of atomic collapse in graphene artificial atoms is not only an effective way to uncover deep mysteries of relativistic atoms and molecules \cite{ZhouNC2024,MaoNature2025}, but also open new horizons to design nanoscale graphene-based devices. 

The breakdown of atomic orbits originates from the fact that the Dirac operator is no longer self-adjoint under a supercritical Coulomb potential \cite{Case, OvdatNC2017}. Therefore, similar atomic collapse phenomenon should also happen in other high-order singular potentials with a general form like $1/r^{\gamma}$ ($\gamma > 1$) \cite{FrankRMP1971,BanePRA2001,BraatenPRA2004}, . Actually, in realistic material systems, the potentials with faster decay characteristic is possible to exist, in view of a non-negligible Coulomb screening effect \cite{RogersPRA1970,HwangPRB2007,ShytovPRL2007B,TerekhovPRL2008}. Theoretical studies have pointed that a nonlinear Coulomb screening could make the strong Coulomb potential asymptotically exhibit the high-order decay at a long range \cite{FoglerPRB2007}, for instance, $1/r^3$ for charged impurities in the doped graphene \cite{KatsnelsonPRB2006}.  Some experiments have also found the modifications of density of states due to the screening effect of potentials \cite{JiangNN2017, LuNC2019}. In addition, high-order decay characteristics of potential fields may also arise in interactions between electrons and electric multipole moments \cite{MartinoPRL2014}.  However, to our knowledge, there is still a lack of theory to universally clarify the behaviors of atomic collapse for high-order singular potentials in graphene, which could be significant to comprehensively understand the uniqueness in relativistic atomic physics.

In this letter, we aim to theoretically clarify the atomic collapse of massless Dirac fermions for a general singular potential $\propto \beta/r^\gamma$  in graphene.  Quite different from the case of Coulomb potential which must demand $\beta > 1/2$, an infinitesimal potential strength $\beta>0$ is shown to be enough to collapse atomic orbits once $\gamma>1$. Moreover, the energy arrangement of ACSs at $\gamma > 1$ is demonstrated to follow a power sequence, rather than the typical geometric-sequence at $\gamma = 1$. Especially, some special ACSs are found to exist even above the bulk Dirac point, which cannot appear in Coulomb potentials. These findings deepen the understanding of unique relativistic properties of massless Dirac fermions in singular potentials and provide some theoretical basis for future investigations on various graphene artificial atoms.

\begin{figure*}[ht]
	\includegraphics[width=0.7\textwidth]{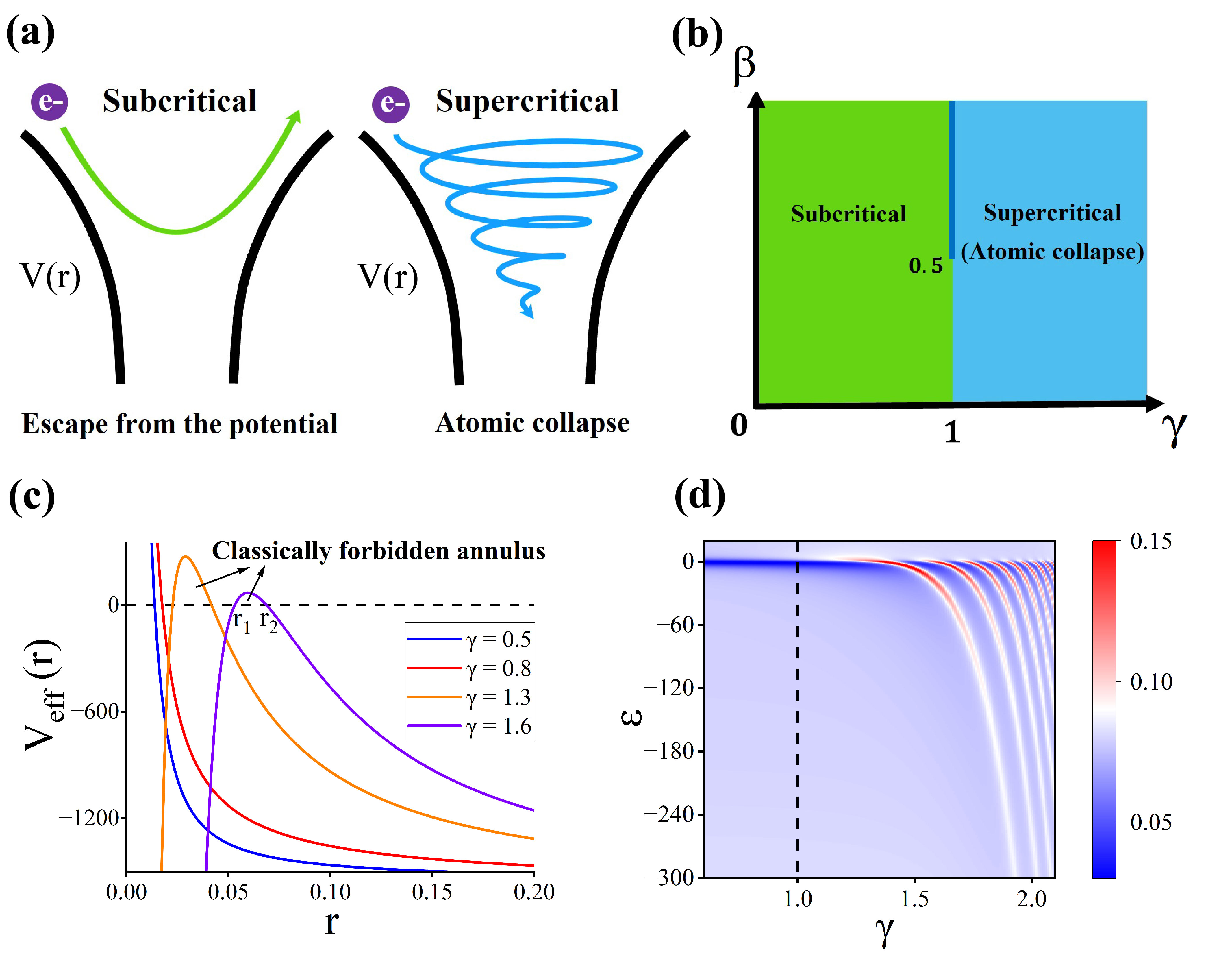}
	\centering %
	\caption{The theoretical analysis of atomic collapse under a general singular potential $\beta/r^{\gamma}$. (a) The schematic diagram of semiclassical electron trajectories in subcritical and supercritical regimes.(b) The phase diagram versus $\beta$ and $\gamma$. At $\gamma =1$, the supercritical regime appears only when $\beta > |m|$ (here for the lowest angular momentum $m=1/2$), which is denoted by a thick blue line. (c) The effective potential $V_{eff}(r)$ for different $\gamma$. Here $\beta = 0.45$, $\epsilon = -40$ and $m = 1/2$. The classically forbidden annulus for $r_1 \leq r \leq r_2$ appears only for $\gamma > 1$. (d) The calculated LDOS map versus $\gamma$ based on the finite difference method with $\beta = 0.45$, where a series of ACS resonant peaks grows once $\gamma > 1$ (dashed black line), indicating an appearance of the supercritical regime.}
	\label{FIG1}
\end{figure*}

\textit{The general analysis and phase diagram}. We consider a massless Dirac equation with an attractive singular potential $V(\mathbf{r})=-(\hbar v_F)^{\gamma}\tilde{\beta}/|\mathbf{r}|^{\gamma}$ \cite{shuoming}:
\begin{equation}
[v_{F}\mathbf{\sigma} \cdot \mathbf{p} + V(\mathbf{r})] \Psi(\mathbf{r}) = E\Psi(\mathbf{r})
\label{EqA}
\end{equation}
with the Fermi velocity $v_{F}$, the Pauli matrix $\mathbf{\sigma}=(\sigma_x, \sigma_y)$, the effective potential strength $\tilde{\beta}>0$, the position vector $\mathbf{r}$ and the kinetic momentum operator $\mathbf{p} = -i \hbar \nabla$. Here $\Psi(\mathbf{r})$ is a two-component spinor and $E$ is energy. Since the system still preserves the rotational symmetry, we can expand the wavefuntion $\Psi(\mathbf{r})$ by using the polar decomposition ansatz: $\Psi_m(|\mathbf{r}|, \theta)=\frac{e^{im\theta}}{\sqrt{|\mathbf{r}|}}\binom{u_1 e^{-i\theta/2}}{iu_2 e^{i\theta/2}}$ with the half-integer angular momentum number $m=\pm \frac{1}{2}, \pm \frac{3}{2}, ...$ Then, Eq.~(\ref{EqA}) can be rewritten as \cite{RodriguezPRB2016}: 
\begin{equation}
	\begin{pmatrix}
		-\frac{\beta}{r^{\gamma}}-\epsilon & \partial_r+\frac{m}{r} \\
		-\partial_r+\frac{m}{r} & -\frac{\beta}{r^{\gamma}}-\epsilon
		\end{pmatrix}\binom{u_1}{u_2}=0.
		\label{EqB} 
\end{equation}
Here dimensionless quantities $r, \epsilon, \beta$ are the reduced radial length $r=|\mathbf{r}|/|\mathbf{r_*}|$, the reduced energy $\epsilon = E/E_*$ and the reduced potential strength $\beta = \tilde{\beta}/\tilde{\beta}_*$. The radial length, energy and potential strength units are chosen as $|\mathbf{r}_*|$, $E_*=\hbar v_F/|\mathbf{r_*}|$ and $\tilde{\beta}_*=(|\mathbf{r}_*|/\hbar v_F)^{\gamma-1}$, respectively. Note that $|\mathbf{r}_{*}|$ can be an arbitrary unit of length here. The behaviors of Dirac quasiparticles can be analyzed by a semiclassical approach based on Wentzel-Kramers-Brillouin (WKB) method \cite{SundaramPRB1999,XiaoRMP2010, Stone, RodriguezPRB2016, HouPRB2019}:
\begin{equation}
p_{r}^2=-V_{eff}(r)=(\frac{\beta}{r^{\gamma}}+\epsilon)^2-\frac{m^2}{r^2}
\label{Eq1}
\end{equation}
where $p_{r}$ is radial momentum. Here we focus on the energy regime $\epsilon \leq 0$ and electrons can move classically only when the effective potential $V_{eff} (r) \leq 0$ (i.e., $p_r^2 \geq 0$). See Fig.~\ref{FIG1}(a), there are two possible scenarios. For a subcritical potential field, $V_{eff} (r \rightarrow 0) \rightarrow + \infty$, relativistic electrons cannot be bounded and can escape away from the potential well [the left panel in Fig.~\ref{FIG1}(a)]. While for a supercritical potential with $V_{eff} (r \rightarrow 0) \rightarrow -\infty $, relativistic electrons can be confined by the potential well and will spiral towards the infinitely deep potential center. At this time, atomic collapse can happen [the right panel in Fig.~\ref{FIG1}(a)]. Specifically, for $\gamma < 1$ or $\gamma > 1$, $V_{eff} (r \rightarrow 0) \rightarrow + \infty$ or $-\infty$ regardless of $\beta$. At $\gamma = 1$ (i.e., Coulomb potential), $V_{eff} (r \rightarrow 0) \rightarrow + \infty$ or $-\infty$ depending on $\beta < |m|$ or $\beta > |m|$. In Fig.~\ref{FIG1}(b), we demonstrate the phase diagram of subcritical (green) and supercritical (blue) regimes versus $\gamma$ and $\beta$. For $0 < \gamma <1$, atomic collapse never happens. For $\gamma = 1$, atomic collapse happens only for $\beta > |m|$ (thick blue line for $m=1/2$). While for $\gamma > 1$, the supercritical regime extends to the whole $\beta > 0$. This finding implies that ACSs may persist even in weak high-order singular potentials or screened Coulomb potentials \cite{KatsnelsonPRB2006}.

Furthermore, Eq.~(\ref{Eq1}) defines a classically forbidden region with $V_{eff} (r) \geq 0$ \cite{ChenPRL2007}, see Fig.~\ref{FIG1}(c). For $0 < \gamma < 1$, the forbidden region is $ 0 \leq r \leq r_1$ with $\beta/r_1^\gamma+\epsilon = |m|/r_1$. The electrons moving towards the potential center will be scattered back by the infinitely high barrier (dark blue and red lines). At $\gamma \geq 1$, this region becomes $ r_1 \leq r \leq r_2$ with $\beta/r_{1,2}^\gamma+\epsilon = \pm |m|/r_{1,2}$, as denoted by orange and purple lines in Fig.~\ref{FIG1}(c). This classically forbidden annulus can trap quantum states within $ 0 < r < r_1$. In scanning tunneling microscopy (STM) experiments, the appearance of atomic collapse can be characterized by resonant peaks in local density of states (LDOS) near the charge center \cite{WangScience2013,MaoNP2016,JiangNN2017,Zheng,MaoNature2025}. To evaluate our conclusions, we use a finite difference method to directly calculate the LDOS map versus $\gamma$ at $r=0$ under a regularized potential $V(r) = - \beta/(r+r_0)^\gamma$ ($r_0$ is the introduced cut-off radius) \cite{RodriguezPRB2016, HouPRB2019}. We set $r_0 = 0.02, \beta = 0.45$ and $m = 1/2$ (see calculation details in Sec.~I in Supplemental Materials \cite{Supp}). Note that here $r_0$ is set as a tunable parameter for the convenience of the subsequent processing and clear comparison with the previous analytic results \cite{ShytovPRL2007}. In principle, the cut-off radius can also be fixed as the length unit (i.e. $r_0 = 1$), and then the derived results are just functions of the reduced parameter $\beta$. In Fig.~\ref{FIG1}(d), we find no resonance peaks when $\gamma<1$. As $\gamma$ climbs to cross $\gamma = 1$, ACSs resonant peaks at the center will gradually grow from the bulk Dirac point, even if $\beta < m$. These results well manifest the phase diagram in Fig.~\ref{FIG1}(b).

\textit{The analysis for the potential $1/r^2$}. Based on the above analysis, we now investigate the properties of ACSs in the high-order singular potential with $\gamma > 1$. As a concise example, we first focus on the case of $\gamma = 2$. Using Einstein-Brillouin-Keller (EBK) quantization rule \cite{SundaramPRB1999,Stone}, the energy of ACSs can be estimated by an integral function:
\begin{equation}
I = \int_{r_0}^{r_1} p_r dr = \int_{r_0}^{r_1} \sqrt{(\frac{\beta}{r^{\gamma}}+\epsilon)^2-\frac{m^2}{r^2}} dr  = n \pi,
\label{Eq2}
\end{equation}
where $n = 1, 2, 3, ...$. It means relativistic electrons move as an enclosed loop $C_{r}$ in the torus space between $r_0$ and $r_1$. For the Coulomb potential, $r_{1,2} = (\beta \mp |m|)/|\epsilon|$. This integral of Eq.~(\ref{Eq2}) can be evaluated with the logarithmic accuracy: $\sqrt{\beta^2-m^2}\ln(\frac{\beta}{r_0 |\epsilon|}) = n \pi$ which gives a geometric series of ACSs \cite{ShytovPRL2007}:
$|\epsilon_{n}|\approx \frac{\beta}{r_0} e^{-\frac{n\pi}{\sqrt{\beta^2 - m^2}}}$. In contrast, for $\gamma = 2$, the classical turning radius is derived as $r_{1,2} = 2\beta/(\pm |m|+ \sqrt{m^2+4\beta|\epsilon|})$. Considering a relatively large energy range $ m^2/\beta \ll  |\epsilon| \ll  \beta/r_0^2 $, $r_{1}$ can be approximated as $r_{1} \approx \sqrt{\beta/|\epsilon|}$. Putting it into Eq.~(\ref{Eq2}) and using the Taylor expansion, the integral $I$ is further approximated as (see derivation details in Sec.~II in \cite{Supp}):
\begin{equation}
I \approx \frac{\beta}{r_0}-2\sqrt{\beta |\epsilon_n|} = n\pi.
\label{Eq3}
\end{equation}
Different from ACSs in the Coulomb potential, the energies of ACSs for $\gamma = 2$ exhibit a power-of-square sequence $|\epsilon_n| \propto  \frac{1}{4\beta}(\frac{\beta}{r_0} - n\pi)^2$.
The deviation from geometric sequences manifests a lack of continuous scale invariance in the system under singular potentials $\propto 1/r^2$. In addition, the energy broadenings due to Klein tunneling of these ACSs  roughly feature the same width-to-energy ratio at a relatively large energy (see details in Sec.~II in \cite{Supp}).

\begin{figure*}[ht]
	\includegraphics[width=0.7\textwidth]{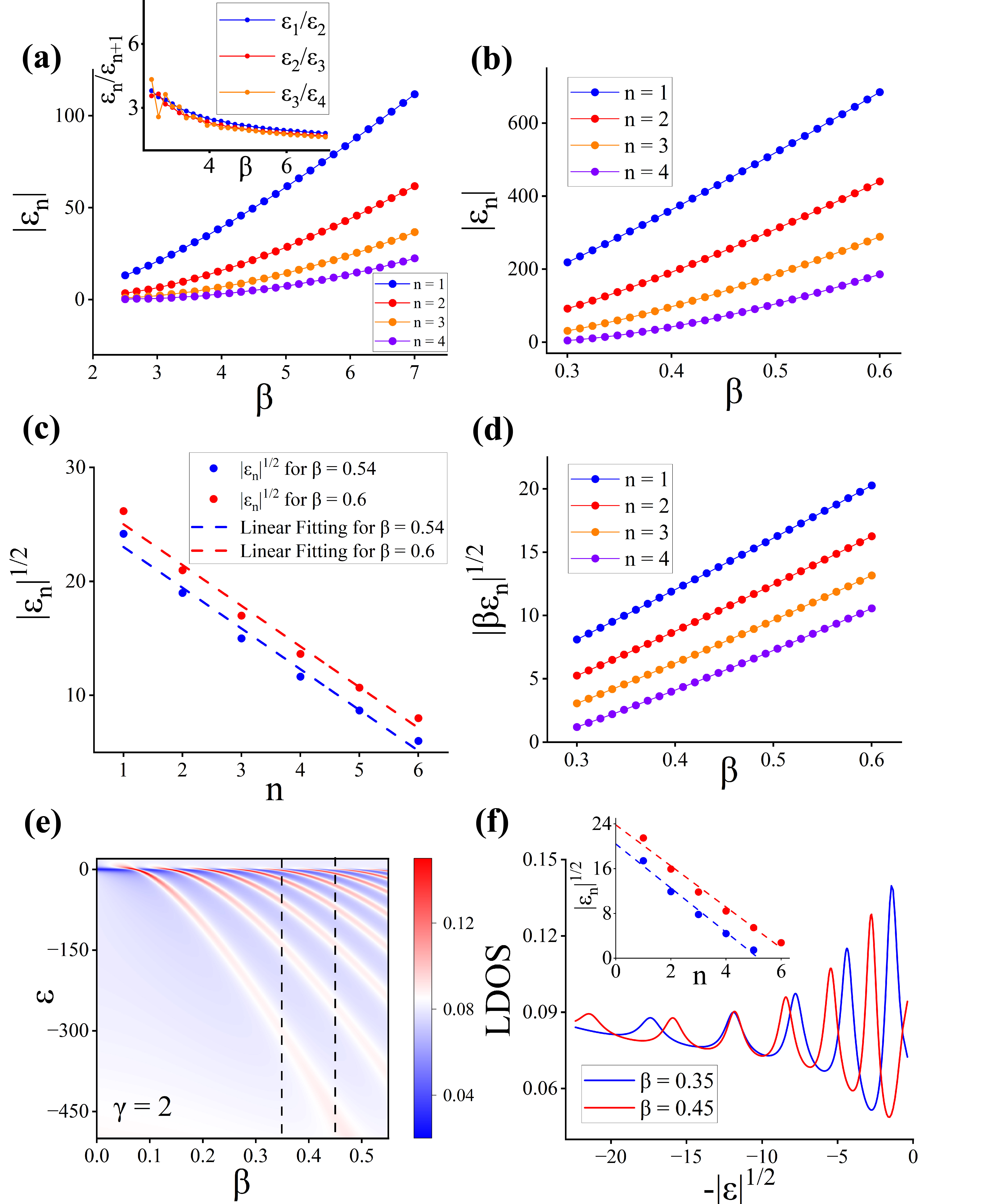}
\centering %
\caption{The behaviors of ACSs at $\gamma=2$. (a) The estimated energies $|\epsilon_n|$ of ACSs from Eq.~(\ref{Eq2}) versus $\beta$ at $\gamma = 1$ for comparison. The inset shows three ratios of adjacent energies $\epsilon_n/\epsilon_{n+1}$ which keep nearly equal. (b) The estimated energies $|\epsilon_n|$ of ACSs versus $\beta$ at $\gamma = 2$. (c) The linear fittings of $|\epsilon_n|^{1/2} \propto  n$ for two different $\beta$ at $\gamma = 2$. (d) The change of $|\beta \epsilon_n|^{1/2}$ versus $\beta$ for the first four ACSs at $\gamma = 2$. (e) The calculated LDOS map versus $\beta$ at the center based on the finite difference method at $\gamma = 2$. (f) The LDOS distributions for two different $\beta$ indicated by black dashed lines in (e). On the axis of $-|\epsilon|^{1/2}$, ACS resonant peaks are roughly arranged as an equal space and still show a linear fitting by $|\epsilon_n|^{1/2} \varpropto n$ (the inset). $m = 1/2$ in the above calculations. }
\label{FIG2}
\end{figure*}

The validity of Eq.~(\ref{Eq3}) can be verified by numerical results in Fig.~\ref{FIG2}. Based on Eq.~(\ref{Eq2}) with a tiny cut-off radius $r_0 = 0.02$ and the lowest angular momentum $|m|=1/2$, the energies $|\epsilon_n|$ of ACSs can be calculated.
For $\gamma=1$, $|\epsilon_n|$ exhibits a nonlinear increasing trend as
$\beta$ increases and the ratios between two adjacent energies $\epsilon_n/\epsilon_{n+1}$ always keep nearly equal which reflects the energy sequences is certainly geometric, see Fig.~\ref{FIG2}(a) and its inset.
However, for $\gamma=2$, $|\epsilon_n|$ deviates from the geometric sequence, although the $\epsilon_n$ still increases with the climb of $\beta$ [Fig.~\ref{FIG2}(b)].
To see it more clearly, in Fig.~\ref{FIG2}(c), we take the $\sqrt{|\epsilon_n|}$ of first six ACSs for two distinct $\beta$ and find they both show good linear fittings with $n$.
In addition, a nearly linear relationship is shown between the $\sqrt{|\beta \epsilon_n|}$ and $\beta$ for each $n$ series, see Fig.~\ref{FIG2}(d). All above results justify the Eq.~(\ref{Eq3}). In Fig.~\ref{FIG2}(e), we also use the finite difference method to numerically calculate the LDOS map versus $\beta$ at $r=0$ based on the regularized potential $V(r)=-\beta/(r+r_0)^2$ with $r_0=0.02$ (see calculation details in Sec.~I in \cite{Supp}). By picking out two distinct $\beta$ (black dashed lines), the resonant peaks of LDOS demonstrate a nearly equally-spaced distribution on the axis of $-|\epsilon|^{1/2}$ [Fig.~\ref{FIG2}(f)]. In the inset, the extracted energies of ACSs exhibit a good fitting of $|\epsilon_n|^{1/2} \propto n$.

\begin{figure*}[ht]
	\includegraphics[width=0.7\textwidth]{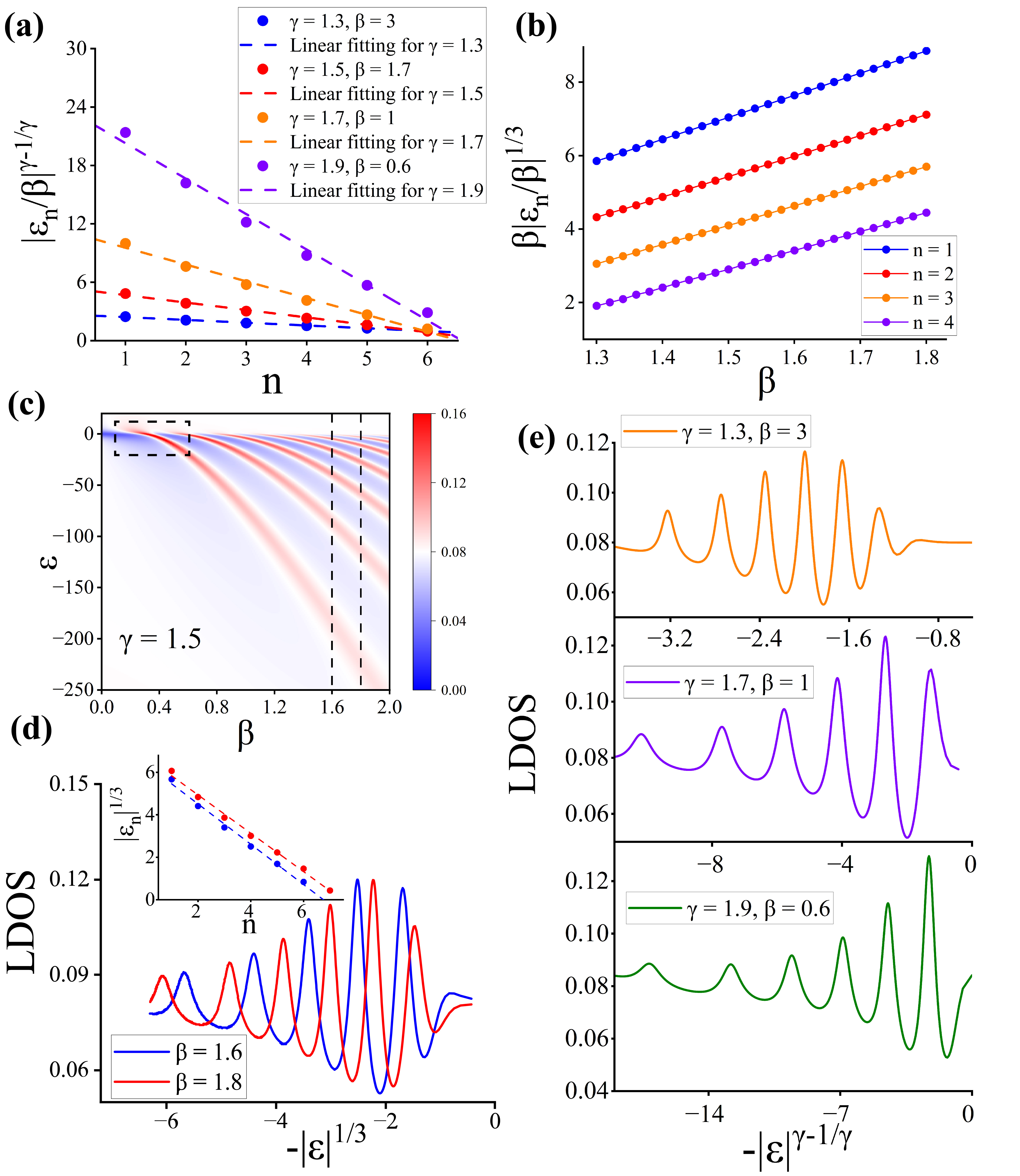}
	\centering %
	\caption{The universal power sequence of ACSs at $\gamma >1$. (a) Linear fittings for estimated $|\epsilon_n/\beta|^{\frac{\gamma-1}{\gamma}} \varpropto n$ at different $\gamma$. (b) The change of $\beta|\epsilon_n/\beta|^{\frac{1}{3}}$ versus $\beta$ for the first four ACSs at $\gamma = 1.5$. (c) The calculated LDOS map versus $\beta$ based on the finite difference method at $\gamma = 1.5$. (d) The LDOS distributions on the axis of $-|\epsilon|^{\frac{1}{3}}$ for two different $\beta$ indicated by black dashed lines in (c). The ACSs resonant peaks are clearly arranged with an equal space, also exhibiting a linear fitting of $|\epsilon_n|^{\frac{1}{3}} \varpropto n$ in the inset. (e) The calculated LDOS distributions on the axis of $-|\epsilon|^{\frac{\gamma-1}{\gamma}}$ based on the finite difference method for several $\gamma$. $m = 1/2$ in the above calculations. }
	\label{FIG3}
\end{figure*}

\textit{The power sequences of ACSs at $\gamma > 1$.} The results shown in Eq.~(\ref{Eq3}) implies that ACSs in inverse power potentials do not follow geometric sequences as ACSs in Coulomb potentials. Actually, they are also unlike another common quasi-bound states, e.g., equally-spaced whispering gallery modes in circular cavities \cite{MatulisPRB2008, Gutierrez2016,ZhaoScience2015}. We wonder whether there is a universal law similar with Eq.~(\ref{Eq3}) for arbitrary $\gamma > 1$. Although the specific forms of $r_{1,2}$ and $\epsilon_n$ are hard to be determined in general cases, some approximations can be adopted. Considering energies $|\epsilon_n|$ of ACSs are usually much larger than $\beta$ and $m$ (as shown in Fig.~\ref{FIG2}), the influence of $m$ can be neglected in solving the classical turning radii, thus $r_{1}$ can be generally approximated as $r_{1} = \sqrt[\gamma]{\beta/|\epsilon|}$ (see details and Fig.~S1 in Sec. III in \cite{Supp}). Putting $r_{1}$ into Eq.~(\ref{Eq2}) and given that $(\beta/r^\gamma + \epsilon)$ plays a leading role in the integrand, we can get:
\begin{equation}
	I \approx \int_{r_0}^{r_1} (\frac{\beta}{r^\gamma}+\epsilon) dr \approx \frac{\beta}{\gamma-1}(\frac{1}{r_0^{\gamma-1}}-\gamma(\frac{|\epsilon_n|}{\beta})^{\frac{\gamma-1}{\gamma}}) = n\pi.
	\label{Eq4}
\end{equation}
Even though the form of Eq.~(\ref{Eq4}) becomes more complicated now, the energies of ACSs still follow a power sequence where the power is determined by $\gamma$. At $\gamma = 2$, Eq.~(\ref{Eq4}) exactly goes back to Eq.~(\ref{Eq3}).

\begin{figure*}[ht]
	\includegraphics[width=0.7\textwidth]{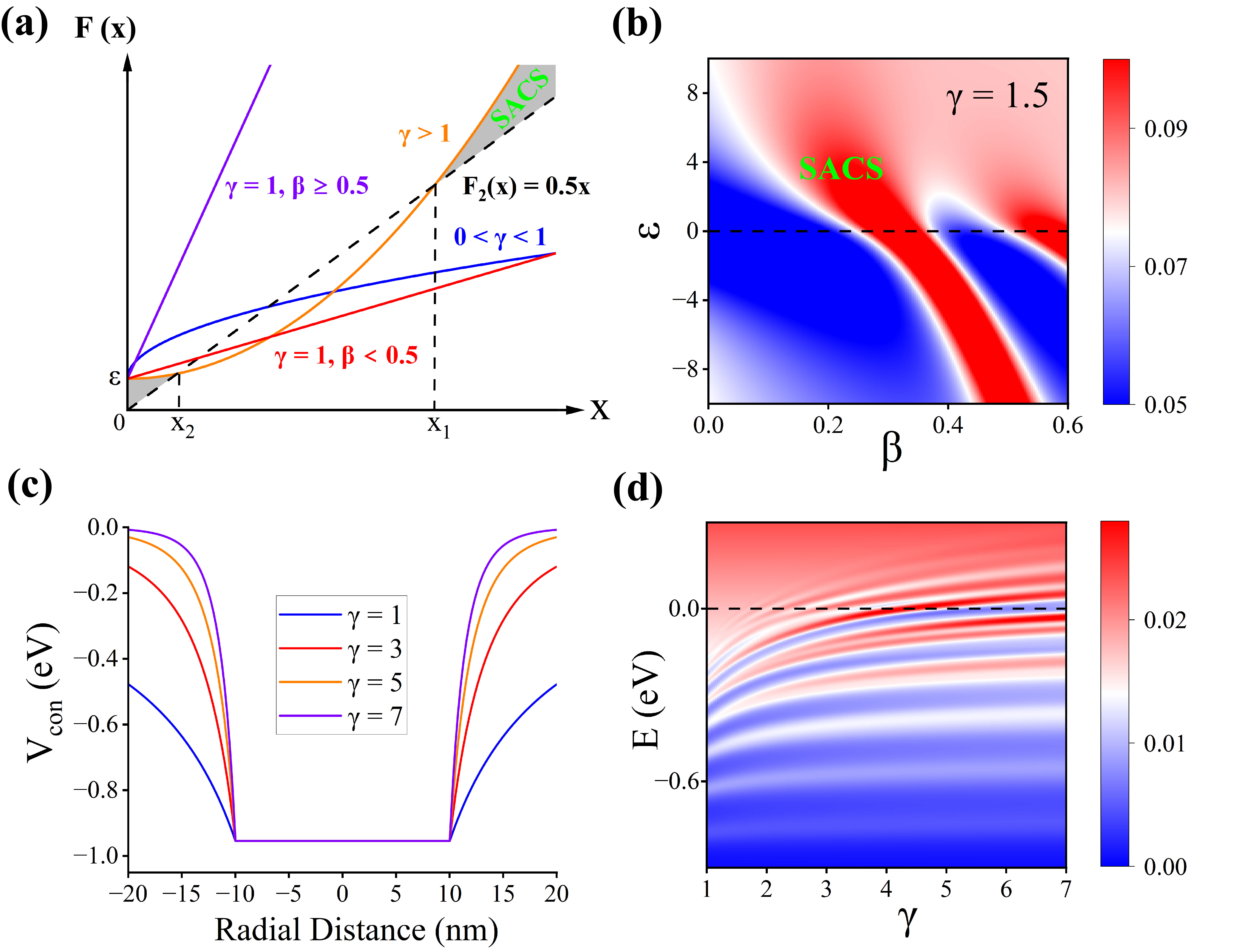}
	\centering %
	\caption{The illustration of SACSs at $\gamma >1$. (a) The WKB analysis to schematically illustrate SACSs denoted by the gray region at $\gamma > 1$. (b) The enlarged figure corresponding to the dashed black box in Fig.~\ref{FIG3}(c). (c) The profiles of $V_{\rm{con}}(\mathbf{r})$ in the unit of eV at different $\gamma$. (d) The simulated LDOS map versus $\gamma$ at the center based on the tight-binding approach. The black dashed line denotes the bulk Dirac point. }
	\label{FIG4}
\end{figure*}

In Fig.~\ref{FIG3}(a), we show the calculated energies of the first six ACSs from Eq.~(\ref{Eq2}) for several $\gamma$. We transform the energy sequence $\epsilon_n$ into $|\epsilon_n/\beta|^{\frac{\gamma-1}{\gamma}}$ and find that each transformed energy sequence exhibits a good linear fitting with $n$. By fixing $\gamma = 1.5$, the transformed energy sequences $\beta|\epsilon_n/\beta|^{1/3}$ also approximately exhibit a linear relationship with $\beta$ for each $n$ series [see Fig.~\ref{FIG3}(b)]. As Eq.~(\ref{Eq4}) indicates, the slopes are roughly the same as the value $1/(\gamma r_0^{\gamma-1}) \approx 4.7$. Based on the finite difference method, we further demonstrate the LDOS of quasi-bound states at the center with a regularized potential $V(r) = -\beta/(r+r_0)^{1.5}$ in Fig.~\ref{FIG3}(c). ACS resonant peaks already emerge when $\beta$ is quite tiny. One abnormal phenomenon is that ACSs exist initially in the positive energy region and gradually dive into the negative region as $\beta$ climbs (denoted by black dashed box). This is against the traditional understanding of ACSs in the Coulomb potential which appear only below the bulk Dirac point \cite{WangScience2013,MaoNP2016,SobolPRB2016}. This point will be discussed later. By picking out two distinct $\beta$ (dashed lines), we can see a series of equally spaced resonances on the axis of $-|\epsilon|^{1/3}$ [see Fig.~\ref{FIG3}(d)] with their extracted energies $|\epsilon_{n}|^{1/3} \propto n$ [see the inset of Fig.~\ref{FIG3}(d)]. Similar results for some other $\gamma$ are together shown in Fig.~\ref{FIG3}(e). Furthermore, we find that the power sequence of ACSs in high-order singular potentials still holds on for higher angular momentum states (see Fig. S2 in Sec.~IV in \cite{Supp}). For more accuracy, we also verify the conclusions on the tight-binding graphene lattice model by an open source code package \cite{pybinding}: $\textit{pybinding}$ via the kernel polynomial method  \cite{Kernel1, Kernel2}  (see more results of Fig.~S3 and Fig.~S4 in Sec.~V in \cite{Supp}). It is also noted in numerical calculations, due to the introduced cut-off, the critical potential strength $\beta_c$ for atomic collapse should be always finite and will be larger as the cut-off radius enlarges \cite{Zeldovich_1972}, but the power-sequence arrangement is robust for different kinds of regularization (See details in Fig. S5 and Fig. S6 in Sec.~VI in \cite{Supp}).

\textit{The special ACSs at $\gamma > 1$.} In Fig.~\ref{FIG3}(c), we find some ACSs above the bulk Dirac point, which is a feature unobserved in a typical Coulomb potential. Actually, a similar phenomenon also appears in the case of $\gamma =2$ in Fig.~\ref{FIG2}(e). These special ACSs (SACSs) can be illustrated from the WKB analysis in Eq.~(\ref{Eq1}). For clarity, we set $x=1/r$ and classical turning radii in Eq.~(\ref{Eq1}) are determined by intersection points of two functions: $F_1(x) = \beta x^\gamma + \epsilon$ and $F_2(x) = |m|x$. Focusing on $|m|=1/2$ and $\epsilon > 0$, the relationship between $F_{1}(x)$ and $F_{2}(x)$ is shown in Fig.~\ref{FIG4}(a). When $0<\gamma<1$, there is only one intersection point between $F_{1}(x)$ (dark blue solid line) and $F_{2}(x)$ (black dashed line). As $x \rightarrow +\infty$ ($r \rightarrow 0$), $F_{1}(x) \ll  F_{2}(x)$ which indicates $V_{eff} (x) \rightarrow +\infty $ and atomic collapse cannot happen. The situation is similar for $\gamma = 1$ and $\beta < 0.5$ (red solid line). For $\gamma = 1$ and $\beta \geq 0.5$ (purple solid line), intersection points disappear and no barriers exist to confine quantum states. For $\gamma > 1$ (orange solid line), $F_1(x)$ changes faster than $F_2(x)$ with two possible intersections $x_{1,2}$ shown in Fig.~\ref{FIG4}(a). Like Fig.~\ref{FIG1}(c), states are trapped by the classical turning radius in the gray region with $x > x_1$ ($0 < r < r_1$). Note that the trapping happens only when $\epsilon < \epsilon_c$. $\epsilon_c$ is the critical point where $F_{1}(x)$ and $F_{2}(x)$ are just tangent:
\begin{equation}
 \epsilon_c =  \frac{|m|^{\frac{\gamma}{\gamma-1}}}{\beta^{\frac{1}{\gamma-1}}}(\gamma^{-\frac{1}{\gamma-1}}-\gamma^{-\frac{\gamma}{\gamma-1}}).
 \label{Eq5}
\end{equation}
Different from ACSs below the Dirac point involving a process of Klein tunneling between electron-like states and hole-like states (similar to positron emission) \cite{KatsnelsonNatP2006,BeenakkerRMP2008, Greiner}, these SACSs with $\epsilon > 0$ exhibit a physical picture about a tunneling only between electron-like states (similar to normal tunneling process). It enriches the atomic collapse physics and demonstrate exotic behaviors of Dirac fermions in high-order singular potentials.

For a fixed $\gamma$, Eq.~(\ref{Eq5}) indicates that $\epsilon_c$ will increase with the decline of $\beta$. See Fig.~\ref{FIG4}(b) which is a enlarged figure corresponding to the dashed box in Fig.~\ref{FIG3}(c), the ACS at $\epsilon < 0$ gradually crosses the bulk Dirac point $\epsilon = 0$ and becomes SACS as $\beta$ decreases. However, the descending of $\beta$ simultaneously reducing the potential strength makes SACS resonant peaks less evident. Therefore, we consider another kind of potentials $V_{\rm{con}}(\mathbf{r})$ in Fig.~\ref{FIG4}(c). $V_{\rm{con}}(\mathbf{r})$ is set as a constant whithin a cut-off radius $|\mathbf{r}_0| = 10$ nm and $V_{\rm{con}}(\mathbf{r}) =  -\hbar v_{F}\beta_0 |\mathbf{r}_0|^{\gamma-1}/\mathbf{|r|}^\gamma$ for $\mathbf{|r|}>|\mathbf{r}_0|$.  By fixing the dimensionless $\beta_0 = 14$, we can find $V_{\rm{con}}$ becomes sharper as $\gamma$ climbs, but the depth of wells $V_{\rm{con}} = -\hbar v_{F}\beta_0/|\mathbf{r}_0|$ remains unchanged [see Fig.~\ref{FIG4}(c)]. This kind of potentials usually appear in graphene quantum dots \cite{Gutierrez2016,Zheng,MaoNature2025}. Next, we construct tight-binding graphene lattice model and simulate LDOS map versus $\gamma$ at the center (see simulation details in Sec.~I in \cite{Supp}). Note that in this model, the radial distance and energy $E$ are in the unit of nm and eV, respectively. In Fig.~\ref{FIG4}(d), a family of ACSs appear only below the Dirac point at $\gamma = 1$. As $\gamma$ rises, ACSs near the Dirac point will move up and finally cross $E = 0$. At $\gamma = 7$, several SACSs have clearly appeared above the Dirac point which is quite distinct from the case of $\gamma = 1$ (also see Fig.~S7 in Sec.~VII in \cite{Supp}). The climb of $\gamma$ slowly turns $V_{\rm{con}}(\mathbf{r})$ into a square potential well, and thus ACSs slowly evolves from power-sequence arrangement to equally-spaced arrangement, as indicated by the power $(\gamma-1)/\gamma \rightarrow 1$ in Eq.~(\ref{Eq4}). The appearance of SACSs can also be a marker to qualitively imply atomic collapse in high-order singular potentials in graphene.

\textit{Discussion and Conclusion}. From the perspective of experimental measurements, ACSs in high-order singular potentials can be directly manifested through STM measurements by characterizing LDOS distributions. Besides, they can be also detected through two-terminal conductance measurements \cite{BardarsonPRL2009}. By connecting the graphene flake to left and right electrodes (grounded) and putting the graphene quantum dot or charge impurity at the center, a two-terminal differential conductance spectrum can be measured by varying the voltage of the left electrode. When the voltage corresponds to position that the incident electron energy aligns with the energy of ACSs, a conductance peak will emerge due to resonant transmissions. However, the two-terminal transport method can only roughly detect energy positions of ACSs but cannot reveal their spatial information.

In summary, we have clarified general features of atomic collapse of high-order singular potentials in graphene. Different from the case in Coulomb potential, a very tiny potential strength is demonstrated to collapse classical atomic orbits and induce a series of ACSs arranged as a power sequence. It is also found that some SACSs can appear even above the bulk Dirac point, which reflects the peculiarity of atomic collapse for massless Dirac fermions at $\gamma > 1$. Our findings will deepen the understanding of atomic collapse in graphene, provide a theoretical basis for future experiments to explore ACSs in diverse artificial atoms and offers new perspectives to construct graphene nanodevices.

\textit{Acknowledgments}. This work was financially supported by the National Natural Science Foundation of China (Grant No. 12447146, No. 12374034 and No. 11921005), the National Key R and D Program of China (Grant No. 2024YFA1409002), the Innovation Program for Quantum Science and Technology (2021ZD0302403), and the Postdoctoral Fellowship Program of CPSF under Grant Number GZB20240031. We also acknowledge the High-performance Computing Platform of Peking University for providing computational resources.

\textit{Data availability.} The data that support this study are available from the authors upon reasonable request.

\bibliography{ref}

\begin{thebibliography}{58}
\bibitem [{shu()}]{shuoming}%
  \BibitemOpen
  \href@noop {} {}\bibinfo {address} {In principle, there are many ways to define the effective potential strength $\tilde{\beta}$ in singular potentials with $\gamma > 1$. The way we define it is to make the question more consise. However, the specific definitions will not affect our conclusions since the main analysis can be dimensionless}\BibitemShut {NoStop}%
\end{thebibliography}

\clearpage
\quad \quad \quad \quad {\large \textbf{Supplemental Materials}}

\section{\label{sec0} SI. The details of numerical calculations}
In this section, we introduce the details of numerical calculations. To obtain the eigenstates and eigenvalues of the massless Dirac radial equation of Eq.~(2) in the main text, we use a finite difference method where $r$ is discretized into $N$ lattices in the interval $ 0 < r < L$ \cite{RodriguezPRB2016, HouPRB2019}. The length is set as $L = 10$ and the lattice number is set as $N = 6000$. To calculate the LDOS $D(\epsilon, m)$ at $r = R_0$ contributed by the $m$ state, we diagonalize the discrete radial Hamiltonian in Eq.~(2) of a fixed $m$ and obtain eigenstates $u_{\alpha, m}(r)$ and eigenvalues $\epsilon_{\alpha, m}$. Then $D(\epsilon,m)$ can be expressed as \cite{RodriguezPRB2016}:
\begin{equation}
D(\epsilon, m) = \sum_{\alpha} \left\langle |u_{\alpha, m} (r = R_0)|^2\right\rangle _{\lambda_d}\delta(\epsilon-\epsilon_{\alpha, m}).
\label{EqS11}
\end{equation}
where the term $\left\langle |u_{\alpha, m} (r = R_0)|^2\right\rangle _{\lambda_d} = \sum_{i} |u_{\alpha, m}(r_i)|^{2}e^{-(r_{i}-R_0)^2/2\lambda_d^2}$ represents a Guassian average of the wavefunction (i denotes the index of discrete latices). In addition, we also introduce a Lorentzian function to simulate the delta function $\delta(\epsilon) \approx \Gamma_d/\pi(\epsilon^2+\Gamma_d^2)$. Here $\lambda_d$ and $\Gamma_d$ denote the spatial broadening and energy broadening, respectively, which reflect the effect of finite-sized STM tips. In the calculations, we set $\lambda_d = 0.2$ and $\Gamma_d = 1$.

In experiments, the potentials in graphene are usually induced by highly charged clusters, vacancies, decorating STM tips and graphene quantum dots \cite{WangScience2013, MaoNP2016,JiangNN2017,Zheng}. To simulate the experiment more realistically, we also construct graphene lattices and use the tight-binding approach to solve the LDOS. The tight-binding Hamiltonian is set as:
\begin{equation}
H_t = \sum_{i}[V(\mathbf{r}_i)+E_{D}]c^{\dagger}_{i}c_{i}-\sum_{\left\langle i,j \right\rangle }tc^{\dagger}_{i}c_{j}.
\label{EqS12}
\end{equation}
Here $i,j$ labels the index of the lattice, $V(\mathbf{r}_i)$ is the potential field at the position $\mathbf{r}_{i}$. $E_{D}$ is the bulk Dirac point which is set as zero in the simulations. The symbol $\left\langle i,j \right\rangle $ denote the nearest-neighbour bonds and $t = 3.2$ eV is the hopping energy which is directly related to the Fermi velocity $v_{F} \approx 1.03 \times 10^6$ m/s by $v_{F} = 3ta_{cc}/2\hbar$ ($a_{cc} = 0.142$ nm is the length of carbon-carbon bond). $c_{i}$ and $c_{i}^{\dagger}$ denote the annihilation and creation operator at site $i$. In tight-binding simulations, we built a large hexagonal graphene flake with all the armchair edges (avoiding zigzag edge-states at the low energy). The side length of the hexagon is 200 nm which is large enough to remove the finite size effect. The numerical calculations are conducted by a open source code package \cite{pybinding}: $\textit{pybinding}$. Based on the kernel polynomial method in this package \cite{Kernel1,Kernel2}, we can solve the LDOS $\tilde{\rho}(E,\mathbf{r}_i)$ at each site $i$ with an appropriate energy broadening $\Gamma_{s}$. Additionally, when scanning the space to obtain space-energy LDOS maps [i.e. Fig.~\ref{FIGS2}(f)], we also introduce an spatial broadening $\lambda_s$ to reformulate the LDOS: $\rho(\mathbf{r}) = \sum_{i} \tilde{\rho}(\mathbf{r_i}) e^{-|\mathbf{r}-\mathbf{r_i}|^2/2\lambda_s^2}$. In the simulations, we set $\lambda_s = 0.15$ nm and $\Gamma_s = 0.01$ eV.

\section{\label{sec1} SII. The derivation for energies of atomic collapse states at $\mathbf{\gamma = 2}$.}

\def\theequation{S\arabic{equation}}
\setcounter{equation}{0}
\def\thefigure{S\arabic{figure}}
\setcounter{figure}{0}
\def\thepage{S\arabic{page}}
\setcounter{page}{1}

To evaluate energies for atomic collapse states within a general singular potentials $\beta/r^\gamma$, we investigate the semiclassical radial equation of WKB method \cite{SundaramPRB1999,XiaoRMP2010,Stone}, as shown in Eq.~(3) in the main text. To estimate the classical forbidden region, we obtain:
\begin{equation}
p^2_r = (\frac{\beta}{r^\gamma}+\epsilon)^2-\frac{m^2}{r^2} = 0
\label{EqS3}
\end{equation}
Here we focus on $\epsilon < 0$ and $\gamma=2$. Using the transformation $r=\frac{1}{x}$, Eq.~(\ref{EqS3}) can be rewritten as:
\begin{equation}
(\beta x^2+\epsilon)^2-m^2x^2=0.
\label{EqS4}
\end{equation}
Then, the classical turning radius $r_{1,2} = \frac{1}{x_{1,2}}$ satisfies:
\begin{equation}
    \beta x_{1,2}^2+\epsilon = \pm |m|x_{1,2}.
    \label{EqS5}
\end{equation}
Since $x_{1,2} > 0$, we can obtain $x_{1,2} = (\pm |m|+\sqrt{m^2+4\beta|\epsilon|})/2\beta$ from Eq.~(\ref{EqS5}).
Assuming that $r_1$ and $r_2$ are respectively the inner radius and outer radius [see Fig.~1(c) in the main text], we have $r_1< r_2$ and $r_1 = 1/x_1 = 2\beta/(|m|+\sqrt{m^2+4\beta|\epsilon|})$. Considering a large energy range $\beta|\epsilon| \gg  m^2$, $r_1$ could be simplified as $r_1 = 1/x_1 \approx \sqrt{\beta/|\epsilon|}$.

Using the EBK quantization rule \cite{Stone}, we further estimate the energies $\epsilon_n$ of quasi-bound states within the potential well by an integral function \cite{RodriguezPRB2016, HouPRB2019}:
\begin{equation}
I = \int_{r_0}^{r_1} p_r dr = \int_{r_0}^{r_1} \sqrt{(\frac{\beta}{r^2}+\epsilon)^2-\frac{m^2}{r^2}} dr.
\label{EqS6}
\end{equation}
The integrand in the right side of Eq.~(\ref{EqS6}) can be further approximated as:
\begin{equation}
	\begin{split}
    \sqrt{(\frac{\beta}{r^2}+\epsilon)^2-\frac{m^2}{r^2}} &= \frac{\beta}{r^2} \sqrt{1+\frac{2\epsilon\beta - m^2}{\beta^2}r^2+\frac{\epsilon^2}{\beta^2}r^4}\\
	&\approx \frac{\beta}{r^2} (1+\frac{\epsilon}{\beta}r^2) + O(r^2).
	\end{split}
    \label{EqS7}
\end{equation}
Here we use a Taylor expansion which is reasonable when $r_{1} \gg  r_0$, namely, $|\epsilon| \ll  \beta/r_0^2$. Considering the integrand in Eq.~(\ref{EqS6}) contributes mainly at $r \rightarrow r_0 \ll  r_1$, and $2 \frac{\epsilon}{\beta} r^2$ should be a small quantity. Putting Eq.~(\ref{EqS7}) into Eq.~(\ref{EqS6}), we can get:
\begin{equation}
	\begin{split}
I &\approx \int_{r_0}^{r_1} \frac{\beta}{r^2} (1+\frac{\epsilon}{\beta}r^2) dr = \beta(\frac{1}{r_0}-\frac{1}{r_1})+\epsilon(r_1-r_0)\\
&\approx \frac{\beta}{r_0} - 2\sqrt{\beta|\epsilon|} = n\pi \qquad (n=1,2,...).
	\end{split}
\label{EqS8}
\end{equation}
where we discard the term $\epsilon r_0$ which is relatively smaller. Therefore, based on appropriate approximations, we can find that atomic collapse states in potentials $\propto 1/r^2$  do not exhibit a geometric sequence as those in the Coulomb potential \cite{ShytovPRL2007,OvdatNC2017}, but a square sequence with $|\epsilon_n| \propto \frac{1}{4\beta}(\frac{\beta}{r_0}-n\pi)^{2}$.

Due to the Klein tunneling \cite{KatsnelsonNatP2006,BeenakkerRMP2008}, the trapped electron-like states can tunnel out of the barrier to become nonlocal hole-like states. This further introduces a finite energy broadening $\Gamma_n \sim  |\epsilon_n|\exp(-2S)$ where $S$ is estimated by \cite{ShytovPRL2007}
\begin{equation}
S \approx \int_{r_1}^{r_2} \mathrm{Im} p_r dr = \int_{r_1}^{r_2} \sqrt{-(\frac{\beta}{r^{\gamma}}+\epsilon)^2+\frac{m^2}{r^2}} dr.
\label{EqS9}
\end{equation}
For the Coulomb potential with $\gamma = 1$, $S \approx \pi (\beta - \sqrt{\beta^2 - m^2})$ which has no energy dependence \cite{ShytovPRL2007}. While at $\gamma =2$, since $\frac{m^2}{r^2}$ is always larger than $(\frac{\beta}{r^2}+\epsilon)^2$ in the classically forbidden annulus $r_{1} \leq r \leq r_2$. $S$ can be roughly estimated as:
\begin{equation}
S \approx \int_{r_1}^{r_2} \frac{|m|}{r} dr = |m| \ln(\frac{r_2}{r_1}) = |m|\ln(\frac{\sqrt{m^2+4\beta|\epsilon|} + |m|}{\sqrt{m^2+4\beta|\epsilon|}-|m|}).
\label{EqS10}
\end{equation}
In the case of $|\epsilon| \gg  m^2/\beta$, $S \approx |m| \ln (\frac{\sqrt{4\beta|\epsilon|}+|m|}{\sqrt{4\beta|\epsilon|}-|m|})$ does not vary significantly with energy $|\epsilon|$. Thus, the series of ACSs in inverse square potentials feature nearly the same width-energy ratio at a relatively large energy.

\section{\label{sec3} SIII.~The derivation of the universal power sequence for atomic collapse states at $\mathbf{\gamma > 1}$.}

\begin{figure*}[ht]
	\includegraphics[width=0.5\textwidth]{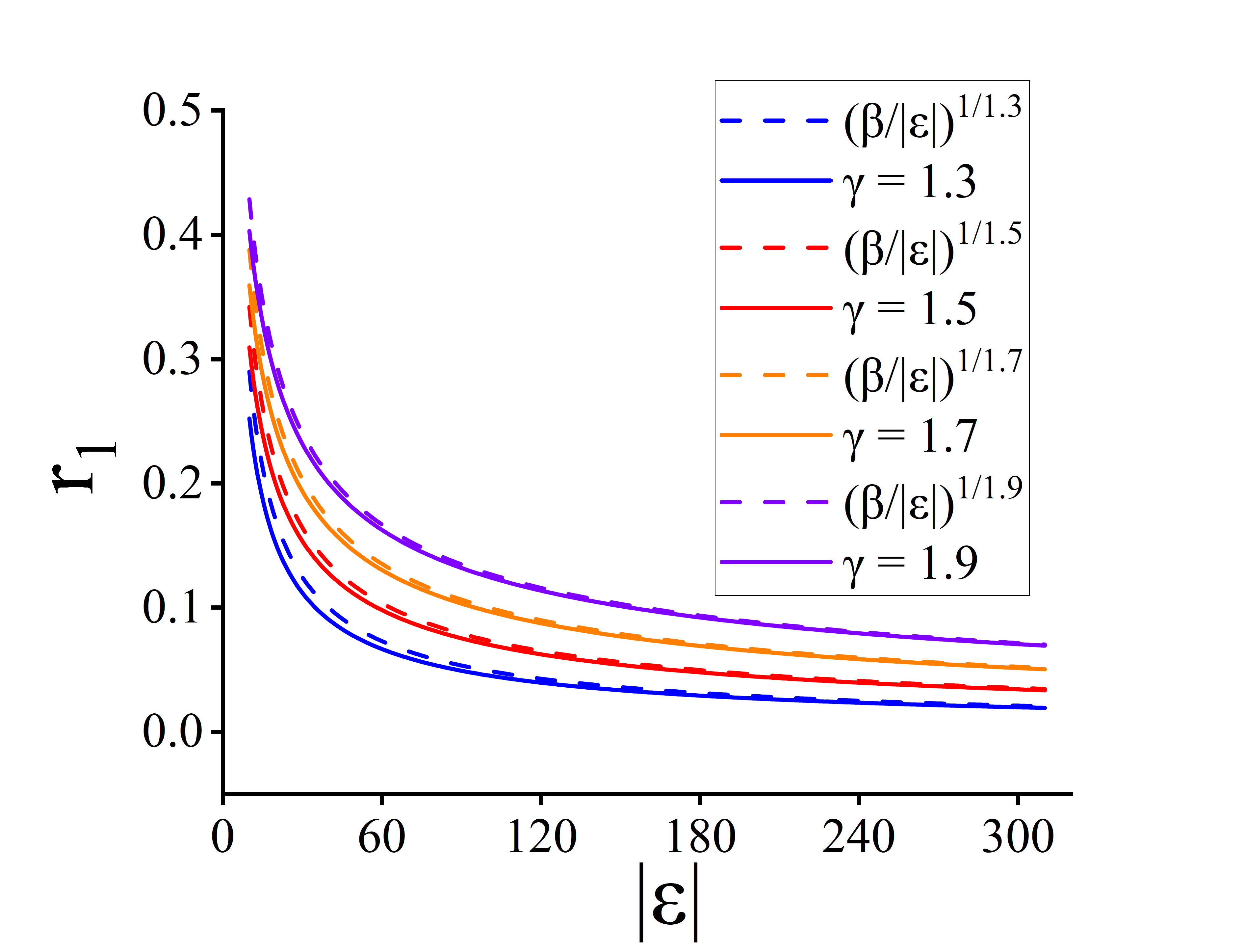}
	\centering %
	\caption{The comparison between the calculated classical turning point $r_{1} = 1/x_1$ from Eq.~(\ref{EqS13}) and the theoretical approximation lines $r_{1}=\sqrt[\gamma]{\beta/|\epsilon|}$ for different $\gamma$. We can find the approximation is valid for a large energy range.}
	\label{FIGS1T}
\end{figure*}

In this section, we will discuss the universal law for energy distributions of ACSs under arbitrary high-order singular potentials ($\gamma > 1$). We first focus on the classical inner turning radius $r_1 = 1/x_1$ by the WKB method:
\begin{equation}
    \beta x_{1}^\gamma + \epsilon - |m|x_1 = 0.
    \label{EqS13}
\end{equation}
Note that here $\epsilon < 0 $. For an arbitary $\gamma >1$, it is difficult to derive a general analytical solution of Eq.~(\ref{EqS13}). But in view of the formulation $x_{1} = \sqrt{|\epsilon|/\beta}$ for $\gamma = 2$, we naturally infer that $x_1$ may share a similar relationship as $x_{1} = \sqrt[\gamma]{|\epsilon|/\beta}$ for the other $\gamma$. Actually, this speculation has some plausibility, because a relatively large $|\epsilon|$ will push $x_1$ towards infinity and $(\beta x_1^\gamma + \epsilon)$ should play a leading role in Eq.~(\ref{EqS13}). To see this more clearly, we assume $x_1 \approx  \sqrt[\gamma]{|\epsilon|/\beta}(1+dx)$ where $dx$ denotes a small deviation. Substituting it into Eq.~(\ref{EqS13}) and using Taylor expansion, we can get $dx \approx \frac{|m|}{\beta}/(-\frac{|m|}{\beta}+\gamma(\frac{|\epsilon|}{\beta})^\frac{\gamma-1}{\gamma})$. Therefore, when $\gamma(\frac{|\epsilon|}{\beta})^\frac{\gamma-1}{\gamma} \gg  \frac{|m|}{\beta}$, $dx$ is indeed a very small quantity and $x_{1} \approx \sqrt[\gamma]{|\epsilon|/\beta}$ should be the approximated solution of Eq.~(\ref{EqS13}). Actually, this approximation holds in most cases in our calculations considering $|m|$ is small and $|\epsilon|$ is usually large. It can be also verified in Fig.~\ref{FIGS1T}, where we demonstrate the calculated $r_{1} = 1/x_1$ versus $|\epsilon|$ from Eq.~(\ref{EqS13}) for four different $\gamma$ with $\beta =2$ (solid colored lines). These calculated $r_{1}$ are found to be well-consistent with theoretical approximation lines $r_{1}=\sqrt[\gamma]{\beta/|\epsilon|}$ (dashed colored lines).

Based on the approximation $r_1 = 1/x_1 \approx \sqrt[\gamma]{\beta/|\epsilon|} $, the energies $\epsilon_{n}$ of ACSs is further estimated using EBK quantization rule:
\begin{equation}
	\begin{split}
I &= \int_{r_0}^{r_1} \sqrt{(\frac{\beta}{r^\gamma}+\epsilon)^2-\frac{m^2}{r^2}}dr \approx \int_{r_0}^{r_1} (\frac{\beta}{r^\gamma}+\epsilon) dr \\
&= \frac{\beta}{\gamma-1}(\frac{1}{r_0^{\gamma-1}}-\frac{1}{r_1^{\gamma-1}}) + \epsilon (r_1-r_0)\\
&\approx \frac{\beta}{\gamma-1}(\frac{1}{r_0^{\gamma-1}}-\gamma(\frac{|\epsilon_n|}{\beta})^{\frac{\gamma-1}{\gamma}}) = n\pi.
	\end{split}
\label{EqS15}
\end{equation}
Similar to the case of $\gamma = 2$, we also use the fact that $\beta/r^\gamma+\epsilon$ is the leading term in the integrand within the trapped region $r_0 \leq r \leq r_1$. We ignore the contribution of $\epsilon r_{0}$ in the last approximation in view of $r_0 \ll  r_{1}$ when $|\epsilon| \ll \beta/r_0^\gamma$.

In conclusion, we analytically deduce that the energies of ACSs in $1/r^{\gamma}$ should follow the power sequence $|\epsilon_n|^{\frac{\gamma-1}{\gamma}} \propto n$ at a relatively large energy. Thus, ACSs in high-order singular potentials are quite distinct from ACSs in Coulomb potentials with geometric sequences. This unique arrangement of energies is a root of the special form of $1/r^\gamma$ ($\gamma > 1$) and the breaking of the continuous scale invariance \cite{OvdatNC2017}. It also helps to characterize the atomic collapse within high-order singular potentials in graphene through STM experiments.

\section{\label{sec4} SIV.~The atomic collapse states in high-order singular potentials with higher angular momenta.}

In the main text, we focus on energies $\epsilon_n$ for ACSs with the lowest angular momentum $m = 1/2$. From the derivations in Sec.~\ref{sec1} and Sec.~\ref{sec3}, the power sequence should also be valid for ACSs even with higher angular momenta.

\begin{figure*}[ht]
	\includegraphics[width=0.9\textwidth]{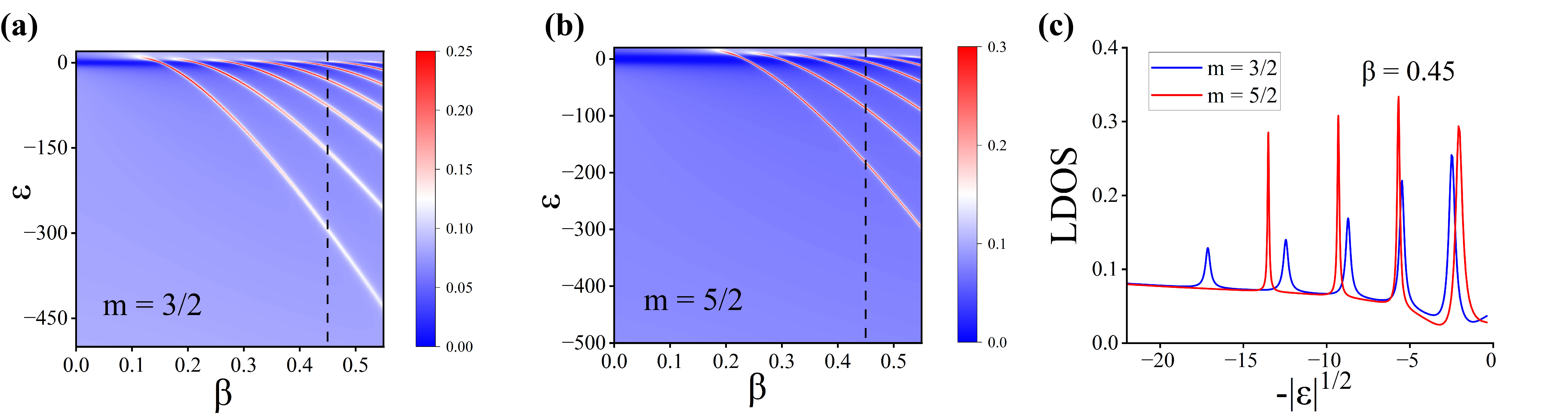}
	\centering %
	\caption{The numerical results for ACSs with higher angular momenta. (a, b) The calculated LDOS map versus $\beta$ at the center for $m = 3/2$ (a) and $m = 5/2$ (b) based on the finite difference method. Here $\gamma = 2$. (c) The extracted LDOS distributions on the axis of $-|\epsilon|^{1/2}$ corresponding to the dashed lines in (a, b) at $\beta = 0.45$.}
	\label{FIGS1}
\end{figure*}

In Figs.~\ref{FIGS1}(a,b), the calculated LDOS map versus $\beta$ with $\gamma = 2$ based on the finite difference method for $m=3/2$ and $m=5/2$, respectively. Compared to Fig.~2(e) in the main text ($m=1/2$), we can find the evolution of ACSs with higher angular momenta show similar trends. However, the emergence of these ACSs requires a relatively stronger coupling strength $\beta$. This is due to the influence of the cutoff radius $r_0$ in $V(r) = -\beta/(r+r_0)^\gamma$. For a bare high-order singular potential $\beta/r^\gamma$ ($\gamma > 1$), the atomic collapse is proved to happen even for a infinitesimal $\beta$, as illustrated by the WKB method and Fig.~1 in the main text. But in real calculations, a finite cutoff radius $r_0$ will impede the emergence of ACSs at $\beta \rightarrow 0$ since the classical turning radius $r_1$ will turn to zero at this time. It is not obvious for ACSs with the lowest angular momentum but is more pronounced for ACSs with higher angular momenta. See $r_1 = 2\beta/(|m|+\sqrt{m^2+4\beta|\epsilon|})$, a larger $|m|$ implies a smaller $r_1$ at the fixed $\beta$ and $\epsilon$.

In Fig.~\ref{FIGS1}(c), we further demonstrate LDOS distributions at $\beta = 0.45$ for $m=3/2$ (dark blue line) and $m=5/2$ (red line), respectively [corresponding to dashed lines in Figs.~\ref{FIGS1}(a,b)]. The specific positions of peaks shift for different angular monmentum states, especially for ACSs with large energies. However, ACSs resonant peaks are still approximately arranged as an equally-spaced sequence on the axis of $-|\epsilon|^{1/2}$, which verifies the validity of our conclusions again.

\section{\label{sec5} SV.~More numerically simulated results on tight-binding graphene lattices.}

In Fig.~2 and 3 in the main text, we verify our theoretical analysis from two aspects: the WKB method and the finite difference method from the massless Dirac Hamiltonian. In order to test our conclusions from a more realistic perspective and provide the guidance for future experiments, we now use the tight-binding approach to simulate the LDOS on graphene lattices and solve the LDOS for energy $E$. 

Similarly, we set a regularized potential field on the graphene lattice: $V(\mathbf{r}_i) = -(\hbar v_{F})^\gamma\frac{\tilde{\beta}}{(|\mathbf{r}_i|+|\mathbf{r}_0|)^\gamma}$. Note that the effective potential strength $\tilde{\beta}$ and energy $E$ in the tight-binding model are not dimensionless quantities. They have units of (eV)$^{1-\gamma}$ and eV, respectively. When we want to make a comparison with the results calculated by the finite difference method, we can nondimensionalize them by $\beta=\tilde{\beta}/\tilde{\beta}_*$ and $\epsilon=E/E_*$ with $\tilde{\beta}_* = (|\mathbf{r}_*|/\hbar v_F)^{\gamma-1}$ and $E_* = \hbar v_F/|\mathbf{r}_*|$. $|\mathbf{r}_*|$ can be estimated by the ratio of the cutoff radius $|\mathbf{r}_0|$ in the tight-binding model and cutoff radius $r_0$ in the finite difference method.

\begin{figure*}[ht]
	\includegraphics[width=0.75\textwidth]{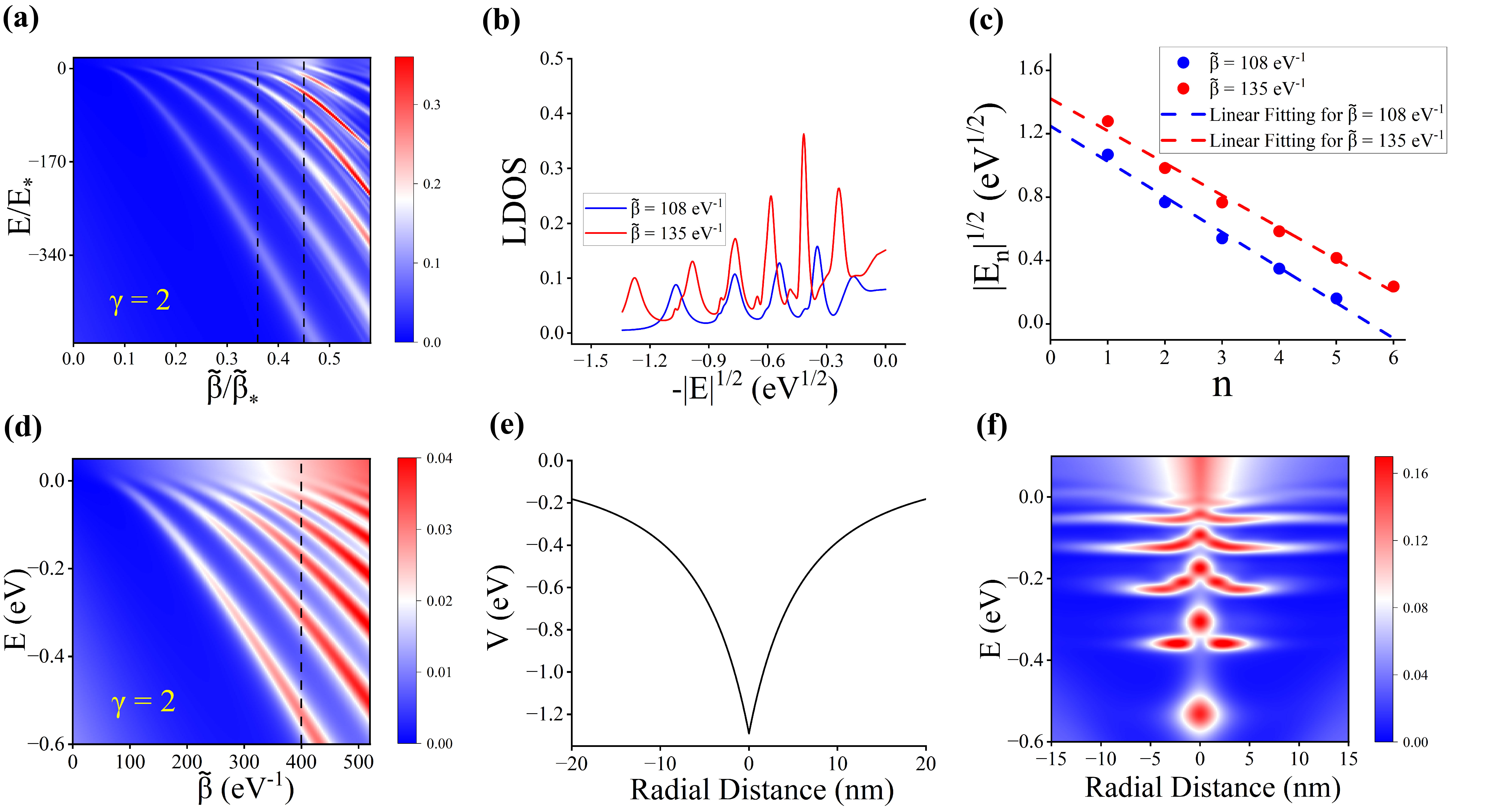}
	\centering %
	\caption{The simulated results on the tight-binding model for $\gamma = 2$. (a) The simulated LDOS map based on the tight-binding method with $|\mathbf{r}_0| = 4$ nm and $\gamma = 2$. For comparison with the Fig.~2(e) in the main text, here we convert the horizontal and vertical coordinates into $\tilde{\beta}/\tilde{\beta}_*$ and $E/E_*$, respectively. (b) The LDOS distributions on the axis of $-|E|^{1/2}$ for $\tilde{\beta} = 108$ eV$^{-1}$ (dark blue line) and $\tilde{\beta} = 135$ eV$^{-1}$ (red line), corresponding to black dashed lines in (a). (c) the extracted ACSs peaks from (b) demonstrate good linear fittings with the index $n$. (d) The simulated LDOS map versus $\tilde{\beta}$ based on the tight-binding method with the same parameters in (a) except for $|\mathbf{r}_0| = 12$ nm. (e) The profile of the potential field $V(\mathbf{r}_i)$ at $\tilde{\beta} = 400$ eV$^{-1}$. (f) The simulated space-energy map of LDOS at $\tilde{\beta} = 400$ eV$^{-1}$ [black dashed line in (d)].}
	\label{FIGS2}
\end{figure*}

In Fig.~\ref{FIGS2}(a), we show the simulated LDOS map versus $\tilde{\beta}$ at the potential center with $|\mathbf{r}_0| = 4$ nm and $\gamma = 2$. To do a comparison with Fig.~2(e) in the main text, we nondimensionalize the horizontal and vertical coordinates by $\beta=\tilde{\beta}/\tilde{\beta}_*$ and $\epsilon=E/E_*$ with $\tilde{\beta}_* \approx 293$ eV$^{-1}$ and $E_* \approx 0.0034$ eV ($|\mathbf{r}_*| = 200$ nm). The simulated result is well consistent with the calculated result shown in Fig.~2(e). It should be pointed out that contributions from all angular momentum states are included in the tight-binding approach, and thus reflect experimental measurements better. Even though there are some splitting peaks which corresponds to higher angular momentum states at a larger $\tilde{\beta}$ (as illustrated in Fig.~\ref{FIGS1}), the major LDOS peaks at the center are still arranged in a power-to-square sequence. See Fig.~\ref{FIGS2}(b), we show extracted LDOS distributions on the axis of $-|E|^{1/2}$ for $\tilde{\beta} = 108$ eV$^{-1}$ and $\tilde{\beta} = 135$ eV$^{-1}$ [black dashed lines in Fig.~\ref{FIGS2}(a)]. There are still good linear fittings between $|E_n|^{1/2}$ of ACSs peaks and index $n$ [Fig.~\ref{FIGS2}(c)].

In a real situation, such a sharp singular potentials will demand a large cutoff radius to avoid a too deep potential fields. Thus, in Fig.~\ref{FIGS2}(d), we consider a large cutoff radius $|\mathbf{r}_0| = 12$ nm and show the simulated LDOS map at the center versus $\tilde{\beta}$. At $\tilde{\beta} \approx 400$ eV$^{-1}$, the depth of the potential well with $\gamma = 2$ is about 1.3 eV (smaller than the hopping energy), as shown in Fig.~\ref{FIGS2}(e). The profile of potential $V(\mathbf{r})$ is somehow similar to the tip-induced potential in Ref. \cite{JiangNN2017}. We can find the evolution of quasibound states still resembles the results shown in Fig. 2(e), and splitting of higher angular momentum states are less evident. This is attributed to spatial deviations from the potential center for ACSs with higher angular momenta, see the space-energy LDOS map at $\tilde{\beta} = 400$ eV$^{-1}$ in Fig.~\ref{FIGS2}(f). In experiments of graphene, the power-sequence ACSs could be more easily detected by the LDOS distributions at the center, especially in a large-sized quantum dots or tip-induced p-n junction \cite{JiangNN2017, Zheng}.

\begin{figure*}[ht]
	\includegraphics[width=0.75\textwidth]{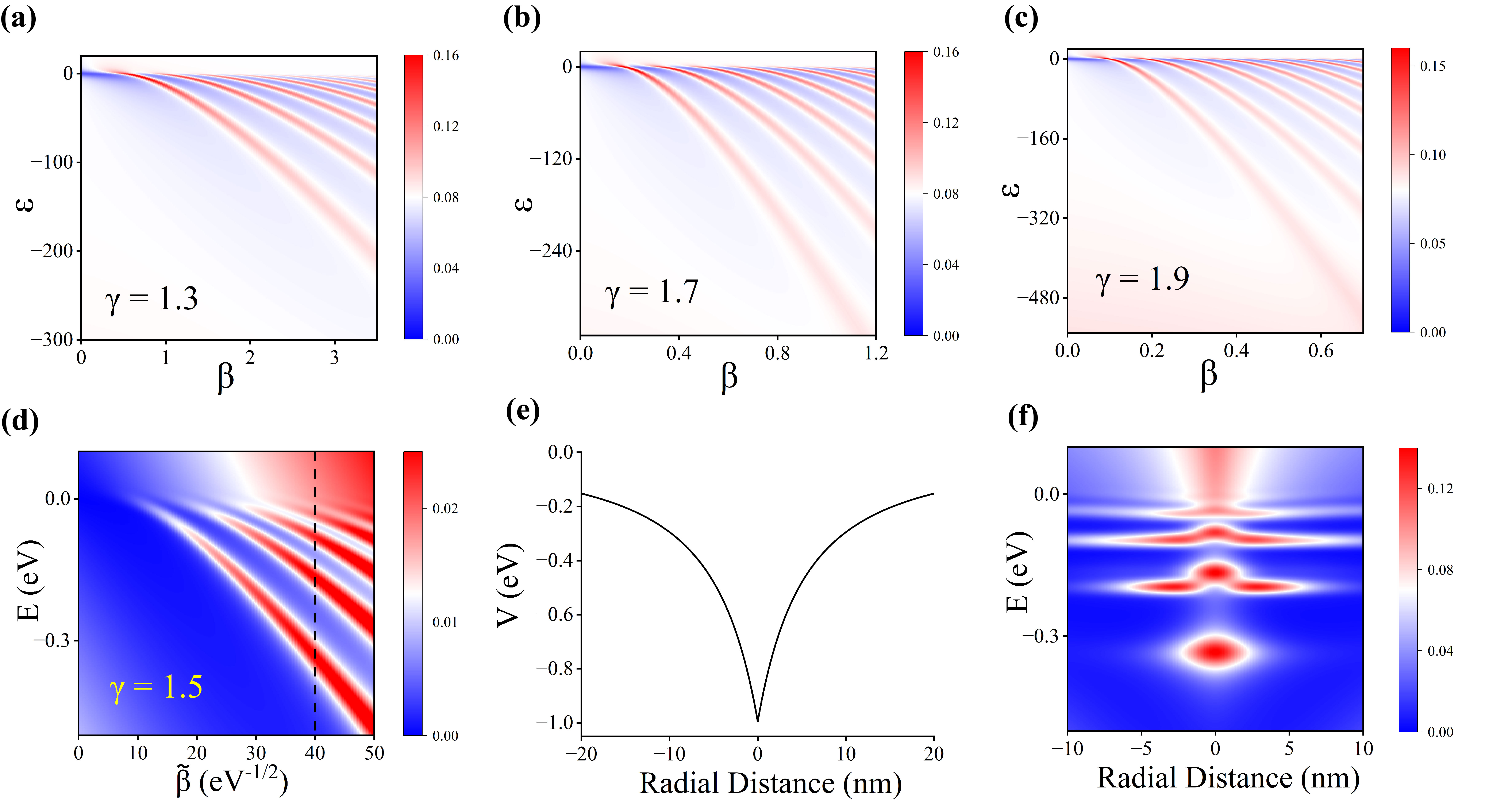}
	\centering %
	\caption{The numerical LDOS maps for different $\mathbf{\gamma}$. (a-c) The calculated LDOS map versus $\beta$ for $\gamma = 1.3, 1.7, 1.9$ based on the finite difference method. (d) The simulated LDOS map versus $\tilde{\beta}$ based on the tight-binding model with $|\bm{r}_0| = 8$ nm and $\gamma = 1.5$. (e) The profile of the potential field $V(\mathbf{r}_i)$ at $\tilde{\beta} = 40$ eV$^{-1/2}$. The simulated space-energy map of LDOS at $\tilde{\beta} = 40$ eV$^{-1/2}$ [black dashed line in (d)].}
	\label{FIGS3}
\end{figure*}

In Fig.~\ref{FIGS3}, we also present numerical results for other $\gamma$. In Figs.~\ref{FIGS3}(a-c), the calculated LDOS maps versus $\beta$ for $\gamma = 1.3, 1.7, 1.9$ based on the finite difference method are shown. They serve as complements to Fig.~3(c) for $\gamma = 1.5$ in the main text. In Fig.~\ref{FIGS3}(d), we show the simulated LDOS map versus $\tilde{\beta}$ at the center based on the tight-binding approach, with $|\mathbf{r}_0| = 8$ nm and $\gamma = 1.5$. The evolution of ACSs is still parallel to the Fig.~3(c). However, due to a large energy broadening, the region for ACSs with $E > 0$ are lesser evident. In Figs.~\ref{FIGS3}(e,f), we also show the profile of potential field $V(\mathbf{r}_i)$ and the simulated space-energy LDOS map at $\tilde{\beta} = 40$ eV$^{-1/2}$ [the dashed line in Fig.~\ref{FIGS3}(d)]. Similar to Fig.~\ref{FIGS2}(f), the central peaks in Fig.~\ref{FIGS3}(f) are mainly related to the lowest angular momentum states ($|m| = 1/2$).

\section{\label{sec6} SVI.~The effect of the potential regularization and the robustness of the power sequence}

\begin{figure*}[ht]
	\includegraphics[width=0.75\textwidth]{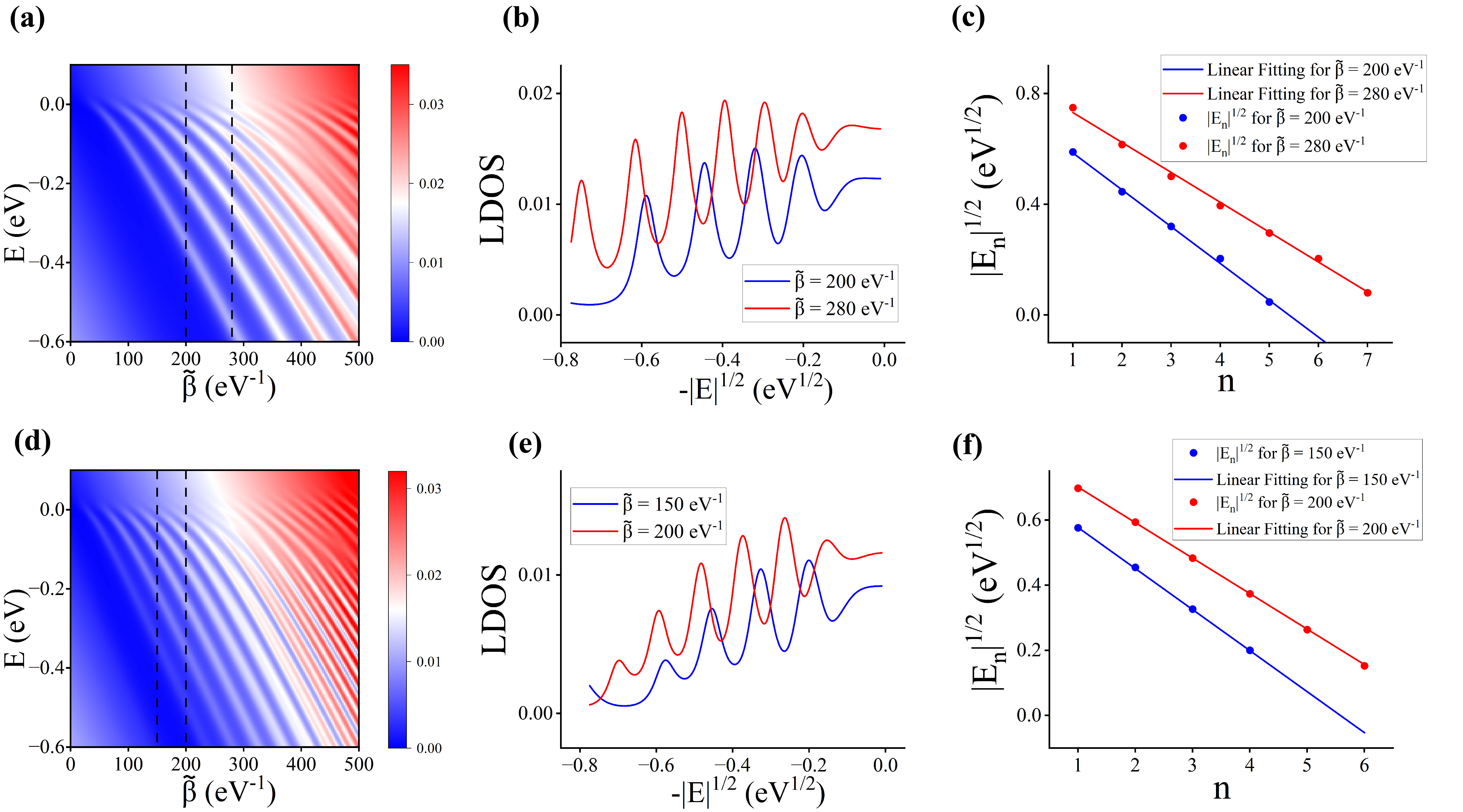}
	\centering %
	\caption{The simulated results for two different kinds of cutoff. (a) The simulated LDOS map versus $\tilde{\beta}$ based on the tight-binding model with potential field $V_{1}(\mathbf{r}_{i})$. (b) The simulated LDOS distributions on the axis of $-|E|^{1/2}$ at $\tilde{\beta} = 200$ eV$^{-1}$, 280 eV$^{-1}$ corresponding to the dashed lines in (a). (c) The extracted ACSs peaks from (b) still demonstrate good linear fittings with the index $n$. (d) The simulated LDOS map versus $\tilde{\beta}$ based on the tight-binding model with potential field $V_{2}(\mathbf{r}_{i})$. (e) The simulated LDOS distributions on the axis of $-|E|^{1/2}$ at $\tilde{\beta} = 150$ eV$^{-1}$, 200 eV$^{-1}$ corresponding to the dashed lines in (d). (f) The extracted ACSs peaks from (e) still demonstrate a good linear fitting with the index $n$.}
	\label{FIGS4}
\end{figure*}

\begin{figure*}[ht]
	\includegraphics[width=0.75\textwidth]{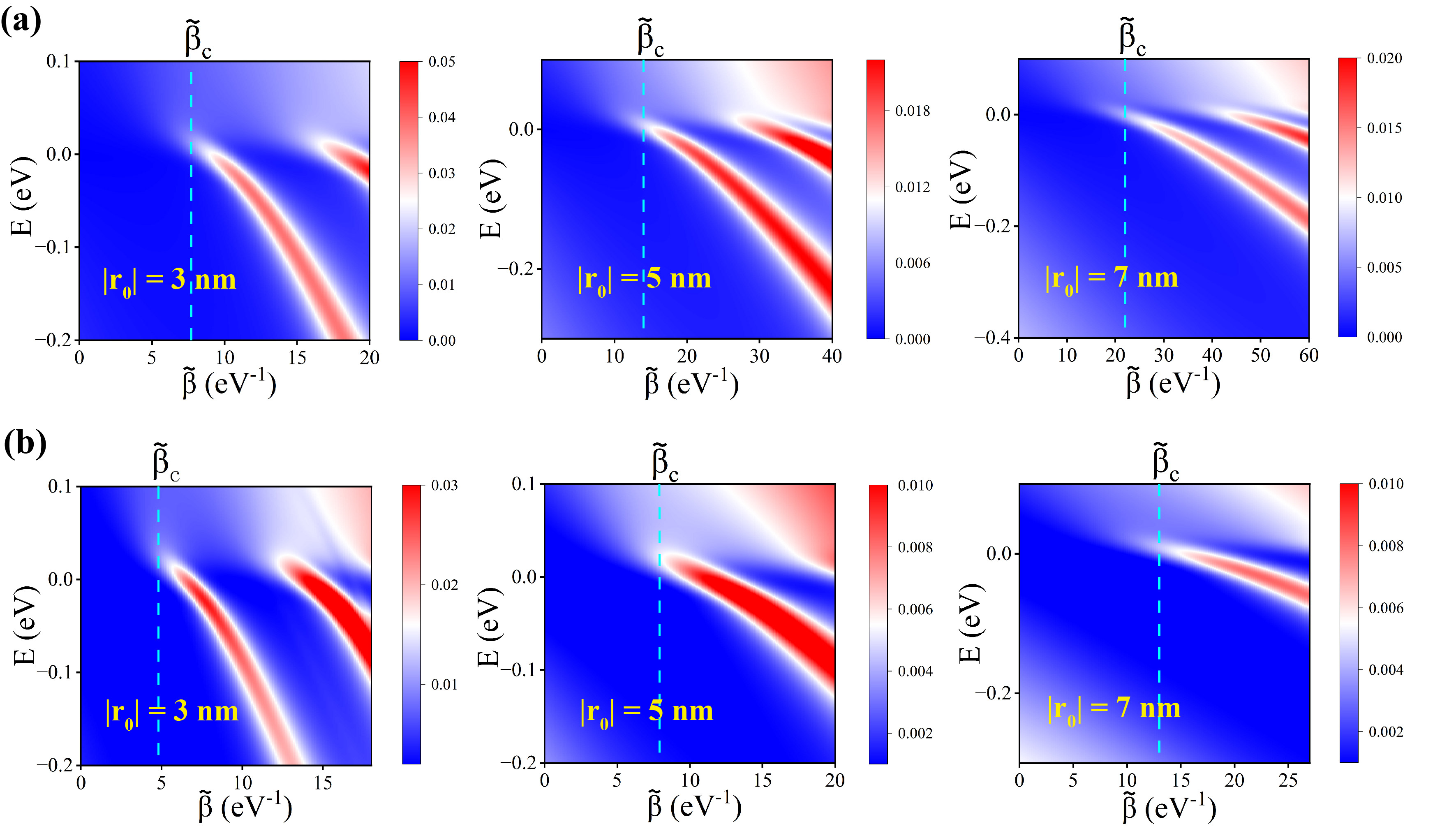}
	\centering %
	\caption{The illustration of the effect for the finite cutoff. (a) and (b) correspond to the evolution process of ACSs with $\tilde{\beta}$ under two different regularized potential fields $V_{1}(\mathbf{r}_i)$ and $V_{2}(\mathbf{r}_i)$, respectively. The cyan dashed lines mark the positions of the critical value $\tilde{\beta}_c$, where the first ACS just emerges.}
	\label{FIGS6}
\end{figure*}

In this section, we will illustrate the effect of the potential regularization and exhibit the robustness of the power-sequence arrangement of ACSs in high-order singular potentials.

In previous calculations, we focus on a specific type of regularized potential field $V(\mathbf{r}) = V_0(\mathbf{r}) = -(\hbar v_F)^{\gamma}\frac{\tilde{\beta}}{(|\mathbf{r}|+|\mathbf{r}_0|)^\gamma}$. Actually, there are many other kinds of cutoff methods. Taking $\gamma = 2$ as an example, we now consider another two regularized potential fields $V_{1}$ and $V_{2}$:
\begin{equation}
	\begin{split}
	V_{1}(\mathbf{r}_i) &= -\frac{(\hbar v_F)^2 \tilde{\beta}}{|\mathbf{r}_i|^2+|\mathbf{r}_0|^2},\\
    V_{2}(\mathbf{r}_{i}) &=
    \left\{
    \begin{aligned}
    &  -\frac{(\hbar v_F)^2 \tilde{\beta}}{|\mathbf{r}_0|^2},  \quad  |\mathbf{r}_i| \leq |\mathbf{r}_0| \\
    &-\frac{(\hbar v_F)^2 \tilde{\beta}}{|\mathbf{r}_i|^2}, \quad |\mathbf{r}_i| > |\mathbf{r}_0| \\
    \end{aligned}
	\right.\\
    \end{split}
    \end{equation}
The different kinds of cutoff methods could correspond to different characteristics of potentials. For example, the form $V_{2}(\mathbf{r}_i)$ with a constant platform implies that the charge distribution in the graphene quantum dot is on the surface, forming an equipotential body \cite{Zheng,Zheng2,ZhouNC2024}. In Figs.~\ref{FIGS4}(a,d), we show the simulated LDOS map versus $\tilde{\beta}$ at the center based on the tight-binding approach for potentials $V_{1}$ and $V_{2}$. Here the cutoff radius $|\mathbf{r}_0|$ is still 12 nm. Compared to Fig.~\ref{FIGS2}(a), the positions and numbers of ACS resonant peaks have changed in the maps. This is reasonable since different types of cutoff will affect the integral in the EBK quantization rule, especially for high-energy ACSs with small $|\mathbf{r}_1|$. However, we emphasize that the approximate law of power-sequences still persists.  See Figs.~\ref{FIGS4}(b,e), the extracted LDOS distributions from Figs.~\ref{FIGS4}(a,d) (see dashed lines) clearly exhibit equal-spaced peaks on the axis of $-|E|^{1/2}$, which still show good linear fittings with the index $n$ [Figs.~\ref{FIGS4}(c,f)].

\begin{figure*}[ht]
	\includegraphics[width=0.75\textwidth]{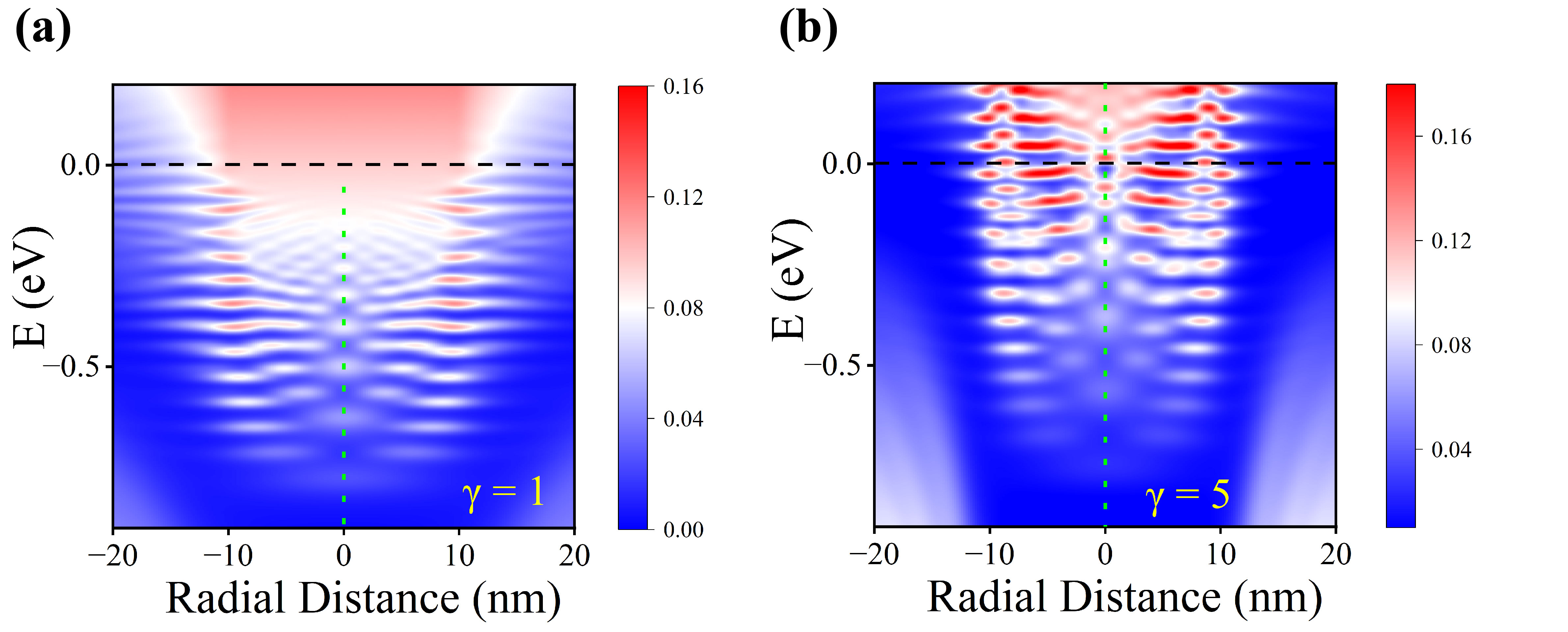}
	\centering %
	\caption{The SACSs in the space-energy maps of LDOS. The simulated space-energy maps of LDOS based on the tight-binding approach for $\gamma = 1$ (a) and $\gamma = 5$ (b) with the other parameters being the same as those in Fig.~4(d). The light green dashed lines denote the ACSs in the center where the SACSs appear above the bulk Dirac point (black dashed lines) at $\gamma = 5$.}
	\label{FIGS7}
\end{figure*}

Although for a standard high-order singular potential field $V(r) \propto  \beta/r^{\gamma} (\gamma > 1)$, we demonstrate that atomic collapse can still appear even for an infinitesimal potential strength $\beta \rightarrow 0$, as shown in Fig. 1. However, the critical value $\beta_c$ required to collapse is actually finite due to the introduction of the regularization. This is because the finite cutoff has erased the singularity at the center. Taking $\gamma =2 $ as an example, when $\beta$ declines towards 0, the classical turning radius $r_1 = 2\beta/(m + \sqrt{m^2 + 4\beta|\epsilon|})$ will simultaneously approach 0. But when $r_1$ shrinks to the range $r_1 \sim r_0$, the WKB picture shown in Fig. 1 has actually been influenced, and it is also hard for massless Dirac fermions be confined in the narrow region $r_0 < r < r_1$ to form ACSs. In addition, as the cutoff radius $r_0$ climbs, the demanding critical value $\beta_c$ should also go up. This characteristic is quite similar to the case that a finite nuclei radius will change the critical charge $Z_c$ from the theoretically predicted value $Z_c \approx 137$ to actually calculated value $Z_c \approx 170$ \cite{Zeldovich_1972}.

To demonstrate this more clearly, we construct the tight-binding graphene lattice model and explore the critical value $\beta_c$ of ACSs for two different regularized potentials $V_1 (\mathbf{r}_i)$ and $V_{2}(\mathbf{r}_i)$ with $\gamma = 2$, as displayed in Figs.~\ref{FIGS6}(a) and (b), respectively. The critical value $\tilde{\beta}_c$ indicating the emergence of the first ACS is roughly denoted by the cyan dashed line. It is shown that $\tilde{\beta}_c$ is always a finite number as the cutoff radius $|\mathbf{r}_0|$ is finite. Furthermore, with the grow of the cutoff radius $|\mathbf{r}_{0}|$, $\tilde{\beta}_c$ is evidently larger, such as from $\tilde{\beta}_c \approx 5$ $\rm{eV^{-1}}$ [Fig.~\ref{FIGS6}(b), $|\mathbf{r}_0| = 3$ nm] to $\tilde{\beta}_c \approx 13$ $\rm{eV^{-1}}$ [Fig.~\ref{FIGS6}(b), $|\mathbf{r}_0| = 7$ nm]. Theoretically, as long as the cutoff radius is very small, $\tilde{\beta}_c$ can approach 0.

\section{\label{sec7} SVII.~The special atomic collapse states at $\mathbf{\gamma > 1}$.}

In Figs.~4(c,d) in the main text, to clearly show the emergence of SACSs, we consider an unique potential configuration $V_{\rm{con}}$:
\begin{equation}
    V_{\rm{con}}(\mathbf{r}_{i}) =
    \left\{
    \begin{aligned}
    &  -\frac{\hbar v_F \beta_0}{|\mathbf{r}_0|},  \qquad  |\mathbf{r}_i| \leq |\mathbf{r}_0| \\
    &-\frac{\hbar v_F \beta_0|\mathbf{r}_0|^{\gamma-1}}{|\mathbf{r}_i|^\gamma}, \qquad |\mathbf{r}_i| > |\mathbf{r}_0|. \\
    \end{aligned}
    \right.
\end{equation}
As shown in Fig.~4(c), the climb of $\gamma$ will continuously sharpen the potential fields, while the depth and the cutoff of $V_{\rm{con}}$ remain constant with $\beta_0 = 14, |\mathbf{r}_0| = 10$ nm. After being nondimensionalized (here choosing $|\mathbf{r}_*| = |\mathbf{r}_0|$ and $\epsilon_*=\hbar v_{F}/|\mathbf{r}_0|$ for simplicity), this potential is reduced as $V_{\rm{con}}=\beta_0/r^\gamma$ when $r>r_0$. Reminding of Eq.~(5) in the main text, we can find $\epsilon_c = |m|(|m|/\beta_0)^{\frac{1}{\gamma-1}}(\gamma^{-\frac{1}{\gamma-1}}-\gamma^{-\frac{\gamma}{\gamma-1}})$ keeps increasing as $\gamma$ climbs ($\gamma > 1$), in view of most of ACSs at the center having an angular momentum $|m| < \beta_0$. Therefore, with the increasing of $\gamma$, some ACSs can gradually cross $E= \epsilon = 0$ and become SACSs in Fig.~4(d) in the main text.

In Fig.~\ref{FIGS7}, we further show the simulated space-energy maps of LDOS corresponding to $\gamma = 1$ (a) and $\gamma = 5$ (b) of Fig.~4(d), respectively. Due to a large cutoff radius $|\mathbf{r}_0| = 10$ nm, a series of quasibound states with high angular momenta are away from the potential center showing the characteristics similar to whispering gallery modes \cite{ZhaoScience2015, Zheng}. For clarity, we focus on ACSs with the lowest angular momentum at the center (denoted by light green dashed lines). At $\gamma = 1$ [Fig.~\ref{FIGS7}(a)], a series of ACSs only extend close to $E = 0$ and the LDOS is continuum beyond the bulk Dirac point (black dashed line). While at $\gamma = 5$, a series of ACSs extend evidently beyond the bulk Dirac point ($E_D = 0$), which corresponds to results shown in Fig.~4(d) in the main text. The existence of SACSs could help to better understand the behaviors of atomic collapse, and also distinct ACSs within higher-order singular potentials.

\end{document}